\documentclass[twocolumn]{aastex62}

\usepackage{subfigure}
\usepackage{tabularx}
\usepackage{multirow}
\usepackage{xcolor}

\received{Septemper 10, 2018}
\revised{December 24, 2018}
\accepted{January 28, 2019}

\submitjournal{ApJ}
\shorttitle{Kim et al. }
\shortauthors{Kim et al.}

\begin{document}

\title{The Synthetic ALMA Multiband Analysis of the Dust Properties \\ of the TW Hya Protoplanetary Disk}


\author{Seongjoong Kim}
\affil{Department of Earth and Planetary Sciences, Tokyo Institute of Technology, 
2-12-1 Ookayama, Meguro-ku, Tokyo 152-8551, Japan}

\author{Hideko Nomura}
\affiliation{Department of Earth and Planetary Sciences, Tokyo Institute of Technology, 
2-12-1 Ookayama, Meguro-ku, Tokyo 152-8551, Japan}

\author{Takashi Tsukagoshi}
\affiliation{National Astronomical Observatory of Japan, 2-21-1 Osawa, Mitaka, Tokyo 181-8588, Japan}

\author{Ryohei Kawabe}
\affiliation{National Astronomical Observatory of Japan, 2-21-1 Osawa, Mitaka, Tokyo 181-8588, Japan}
\affiliation{The Graduate University for Advanced Studies (SOKENDAI), 2-21-1 Osawa, Mitaka, Tokyo 181-0015, Japan}
\affiliation{Department of Astronomy, School of Science, University of Tokyo, Bunkyo, Tokyo, 113-0033, Japan}

\author{Takayuki Muto}
\affiliation{ Division of Liberal Arts, Kogakuin University, 1-24-2 Nishi-Shinjuku, Shinjuku-ku, Tokyo 163-8677, Japan}

\begin{abstract}

Analyzing multiband observations of dust continuum emission is one of the useful tools to constrain dust properties which help us to understand the physical properties of the disks. We perform the synthetic ALMA multiband analysis to find the best ALMA band set for constraining the dust properties of the TW Hya protoplanetary disk. We find that the Band [10,6,3] set is the best set among the possible combinations of ALMA Band [3,4,5,6,7,8,9,10]. We also find two conditions for the good ALMA band sets providing narrow constraint ranges on dust properties; (1) Band 9 or 10 is included in the band set and (2)  Enough frequency intervals between the bands. These are related with the conditions which give good constraints on dust properties: the combination of optically thick and thin bands are required, and large $\beta$ ($\beta$ is the power-law index of dust opacity, $\kappa_{\nu}\propto\nu^{\beta}$) and low dust temperature are preferable. To examine our synthetic analysis results, we apply the multiband analysis to ALMA archival data of the TW Hya disk at Band 4, 6, 7, and 9. Band [9,6,4] set provides the dust properties close to the model profile, while Band [7,6,4] set gives the dust properties deviating from the model at all radii with too broad constraint range to specify the accurate values of dust temperature, optical depth, and $\beta$. Since these features are expected by the synthetic multiband analysis, we confirm that the synthetic multiband analysis is well consistent with the results derived from real data.

\end{abstract}


\section{INTRODUCTION}
\label{sec:introduction}

With the high-resolution ALMA observation of HL Tau \citep{ALMAPartnership2015}, TW Hya has been one of the representative protoplanetary disks (PPDs) which have axisymmetric ringed structure. The central star of TW Hya is a T Tauri star with the mass of $\sim 0.8~M_\odot$ \citep{Andrews2012}. The gas-rich protoplanetary disk surrounding TW Hya has the disk mass $M_{disk}>0.05~M_{\odot}$ derived from HD emission line observations by Herschel. The estimated age of the disk is $\sim~3-10~Myr$ \citep{Bergin2013}. Considering the typical lifetime of gaseous PPDs is $\sim3~Myr$ \citep{Hernandez2007}, the TW Hya disk is an old population. Its proximity, $d\sim59\pm1~pc$ \citep{GAIA2016}, provides good chance to see the detailed structure of the disk ($0.1''$ corresponding $\sim6~AU$). By those characteristics, the TW Hya disk is one of the important targets to study the formation of axisymmetric gap structures and the planet formation in the disks.

Recently, ALMA high spatial resolution observations have confirmed its clear multiple ring structure \citep{Tsukagoshi2016,Andrews2016,Huang2018}. This structure has been thought to be an important signpost of planet formation in the disks. Although some theoretical/numerical scenarios have been established for an individual object to explain the origins of their disk structures, there is no clear general consensus yet; (1) gravitational interaction between a protoplanet and the disk \citep[e.g.][]{Kanagawa2016,Dong2017,Bae2017,Fedele2018}, (2) Magneto-rotational instability \citep[e.g.][]{Flock2015}, (3) secular gravitational instability \citep[e.g.][]{Youdin2011,Takahashi2014}, and (4) growth and destruction of dust particles near snowlines \citep[e.g.][]{Zhang2015,Okuzumi2016}. The origin of gap structures in the TW Hya disk is also still under debate.

Dust properties such as dust temperature $T_{d}$, dust opacity $\kappa_{\nu}$, and dust opacity power-law index $\beta$ are the key parameters to distinguish the models explaining the origin of those disk structures. They are expected to behave differently in the disk structures like rings, gaps or spiral arms depending on the formation mechanism of the structures. 
\citet{Beckwith1990} and \citet{Beckwith1991} are the pioneering works to constrain $\tau_{\nu}$ and $\beta$ from the millimeter wavelength single-dish observations of the pre-main sequence stars in Taurus-Auriga clouds. \citet{Kitamura2002} extended this approach to the interferometer observations to additionally measure the outer radii of the disks and then estimate the dust properties. After that, some studies \citep[e.g.][]{Andrews2007,Andrews2009} made further constraints on the dust properties by comparing the theoretical models with the observed visibility data. Moreover, lots of studies \citep[e.g.][]{Guilloteau2011,Perez2012,Perez2015,Tazzari2016} have tried to derive the radial profiles of dust properties, such as $\tau_{\nu}$ or $\beta$, from the interferometric observations of various PPDs. In those works, the radial profiles of dust temperature have been mainly derived using the SED modeling and the radiative transfer calculations or a simple assumption of dust temperature profile is adopted. But the ALMA observations with high spatial resolution have made it possible to derive the temperature profiles directly from the observed intensity profiles.

As a previous research, \citet{Tsukagoshi2016} also derived the radial profiles of $\tau_{\nu}$ and $\beta$ from the dust continuum observations of the TW Hya disk at ALMA Band 4 and 6 with an assumed dust temperature radial profile. They made meaningful constraints on the dust properties in the disk from the deprojected azimuthally-averaged intensity profiles of dust continuum emission obtained by the high-resolution ALMA observations. However, since the uncertainty of dust size distribution and then the dust opacity, and also the structures in the disk, such as gaps opened by the planets, will affect the dust temperature profile \citep{Jang-Condell2012}, more direct observations are necessary for better understanding of the dust properties without any assumption of the dust temperature. In order to drop the assumed dust temperature profile, and then obtain the improved radial profiles of dust properties, we need additional observations at a different frequency. 

In this paper, we perform the synthetic ALMA multiband analysis to find which ALMA band set provides us good constraints on the dust properties in the TW Hya disk. And then, we apply this multiband analysis to ALMA archival data to examine whether the results derived from ALMA archival data of the TW Hya disk are consistent with our synthetic analysis. In Section \ref{sec:methodology}, we describe how the synthetic ALMA multiband analysis derives the dust properties. Section \ref{sec:results} describes the results of synthetic multiband analysis and the conditions required for good band sets. We will explain the dust properties derived from ALMA archival data through multiband analysis and the consistency between the synthetic results and real data in Section \ref{sec:application}. Section \ref{sec:discussion} will discuss some points to improve the multiband analysis. Finally, Section \ref{sec:Summary} summarizes the conclusions of our multiband analysis.

\section{The Synthetic ALMA Multiband Analysis}
\label{sec:methodology}

For finding which ALMA band set provides us good constraints on the dust properties of the TW Hya disk, we perform the synthetic ALMA multiband analysis. This synthetic analysis consists of two parts, setting up the model radial profiles of intensity at each ALMA band and performing sensitivity analysis for extracting the constraint ranges of synthetic dust properties corresponding to the model intensity and observational errors. We apply the synthetic multiband analysis for all the possible combinations of three ALMA bands among ALMA Band 3 to Band 10. Table 1 lists the representative frequency and frequency coverage of each ALMA band used in the synthetic multiband analysis in a $GHz$ unit.
					
\begin{table}[t]					
\begin{center}
\label{Table1}											
\caption{The representative frequency and frequency coverage of each ALMA Band used in the synthetic multiband analysis}
\begin{tabular*}{0.45\textwidth}{@{\extracolsep{\fill}} l | c c c c c c c c }			
\hline \hline					
Band	  &	3	&	4 & 5 & 6 & 7 & 8 & 9 & 10	\\
\hline					
$\nu_{min}$ [GHz]	&	84 & 125 & 166 & 211 & 275 & 385 & 602 & 787	\\
\hline
$\nu_{rep}$ [GHz]	&	112 & 145 & 190 & 233 & 346 & 450 & 670 & 870	\\
\hline
$\nu_{max}$ [GHz]	&	119 & 163 & 211 & 275 & 370 & 500 & 720 & 950	\\
\hline \hline					
\end{tabular*}	
\end{center}					
\end{table}

\subsection{The model radial intensity profiles at each ALMA band}
\label{subsec:model}

As a first part of the synthetic multiband analysis, we calculate the model radial intensity profiles at each ALMA band. For the calculation, we use the formal solution of the radiative transfer equation for dust continuum emission, 
\begin{equation}
    I_{\nu}(r) = B_{\nu}(T_{d}(r))[1-exp(-\tau_{\nu}(r))]
\end{equation}
where $B_{\nu}(T_{d})$ is the Planck function at dust temperature $T_{d}$ and frequency $\nu$. $\tau_\nu$ is the dust optical depth for which we adopt 
\begin{equation}
    \tau_{\nu}=\tau_0(\nu/\nu_0)^{\beta}
\end{equation}
by assuming the dust opacity obeys the power-law, $\kappa_{\nu}\propto \nu^{\beta}$. In this case, the spectral index $\alpha$ of dust continuum emission can be written as
\begin{equation}
    \alpha (r) = 3 - \frac{h\nu}{k_{B} T_{d}(r)} \frac{e^{h\nu/k_{B}T_{d}(r)}}{e^{h\nu/k_{B}T_{d}(r)} - 1} + \beta\frac{\tau_{\nu}(r)}{e^{\tau_{\nu}(r)} - 1}
\end{equation}
where $k_{B}$ Boltzmann constant and $h$ Planck constant. We note that we simply assume that the source function is $S_{\nu}=B_{\nu}(T_{d})$, and the dust temperature $T_{d}$ and the absorption coefficient, $\kappa_{\nu}\rho_{d}$ ($\rho_{d}$ is the dust density), are constant along the line-of-sight in equation (1) for simplicity.

First, we derive the frequency independent radial profiles $\tau_{0,model}(r)$ and $\beta_{model}(r)$ using Equation (1)$-$(3) and the observed radial intensity profiles, $I_{obs}(r)$, and spectral index, $\alpha_{obs}(r)$, at $190~GHz$ together with the assumption of $T_{d, model}(r)=26K(r/10AU)^{-0.4}$. The observations were performed with high-resolution ($\sim88.1$ mas$\times$62.1 mas) by ALMA \citep{Tsukagoshi2016}. The $T_{d,model}(r)$ profile is obtained by fitting the temperature profile at the disk midplane in \citet{Andrews2012,Andrews2016}, which is derived using radiative transfer calculations. We note that the distance of TW Hya is assumed as $d\sim54~pc$ in the previous works, but we used the updated value $d\sim59~pc$ \citep{GAIA2016}. Using the derived $\tau_{0, model}(r)$ and $\beta_{model}(r)$, we calculate frequency dependent optical depth $\tau_{\nu,model}(r)$ from equation (2). Then we obtain the model radial intensity profiles, $I_{\nu,model}(r)$, at each ALMA band using Equation (1) with the assumed $T_{d, model}(r)$ and the derived $\tau_{\nu, model}(r)$. 

\subsection{Sensitivity analysis}
\label{subsec:sensitivity}

After calculating the model radial intensity profiles and frequency-independent $\tau_{0,model}$ and $\beta_{model}$, we perform the sensitivity analysis to extract the constraint ranges of the synthetic dust temperature $T_{d,syn}$, optical depth $\tau_{\nu,syn}$, and opacity power-law index $\beta_{syn}$ at a given radius. To perform this analysis, we establish the parameter space of synthetic intensity $I_{\nu,syn}$ at three frequencies which are picked arbitrarily among ALMA Band 3 to 10 corresponding to $T_{d,syn}$, $\tau_{\nu,syn}$, and $\beta_{syn}$. Then, we extract the constraint ranges of synthetic dust properties satisfying the ranges of $(100\%-x)I_{\nu,model}\leq I_{\nu,syn}\leq (100\%+x)I_{\nu,model}$ where $I_{\nu,model}$ is the model radial intensity profile derived in Section \ref{subsec:model} and $x$ are observational errors which vary depending on ALMA band. 

We practically obtain the constraint ranges of the synthetic dust properties $T_{d, syn}$, $\tau_{\nu, syn}$ and $\beta_{syn}$ in the following way. We pick three different ALMA bands [$\nu_{1}, \nu_{2},\nu_{3}$] among ALMA Band [3,4,5,6,7,8,9,10] as one band set. Then, at a given radius, we set $I_{\nu,syn}$ which corresponds to the range of $(100\%-x)I_{\nu, model}\leq I_{\nu, syn}\leq(100\%+x)I_{\nu, model}$ for $\nu=\nu_{1}$ and $\nu_{2}$ where the accuracy of amplitude calibration for ALMA data $x=10\%$ for Band 3, 4, 5 and 6, $x=15\%$ for Band 7 and 8, $x=20\%$ for Band 9 and 10 \citep[][p.155]{ALMATech2018}. Next, we derive $\tau_{\nu_{1},syn}$, $\tau_{\nu_{2},syn}$ and $\beta_{syn}$ from $I_{\nu_{1},syn}$, $I_{\nu_{2},syn}$ and Equations (1)-(2) with the synthetic dust temperature $T_{d,syn}=5-60~K$. And then, we estimate $I_{\nu_{3},syn}$ using $T_{d,syn}$, $\tau_{\nu_{1},syn}$, $\tau_{\nu_{2},syn}$, $\beta_{syn}$ and Equations (1)-(2). Finally, we extract the sets of $[T_{d,syn},\tau_{\nu,syn},\beta_{syn}]$ which satisfy $(100\%-x)I_{\nu_3, model}\leq I_{\nu_3, syn}\leq(100\%+x)I_{\nu_3, model}$. Performing these calculations for every radius, we obtain the minimum and maximum values of $T_{d,syn}$, $\tau_{\nu,syn}$ and $\beta_{syn}$, which corresponds to $(100\%-x)I_{\nu,model}\leq I_{\nu,syn}\leq (100\%+x)I_{\nu,model}$ at three frequencies $\nu=[\nu_{1},\nu_{2},\nu_{3}]$ as shown in Figure 2.

\section{The results of the synthetic ALMA multiband analysis}
\label{sec:results}

\subsection{The synthetic multiband analysis results}
\label{subsec:synthetic_results}

We perform the synthetic ALMA multiband analysis (see Section \ref{sec:methodology}) for all the possible combinations of three bands among  ALMA Band 3 to 10. Among total 56 possible combinations, some ALMA band sets provide us narrow constraint ranges on the radial profiles of $T_{d,syn}$, $\tau_{\nu,syn}$, and $\beta_{syn}$, while the others have very broad constraint ranges of the synthetic dust properties. Since the upper limit of the synthetic dust temperature ($T_{d,syn,max}$) is sensitive to the observational errors in the synthetic intensities between $20~AU$ and $45~ AU$, as shown in Figure 2, we calculate the average normalized deviation of $T_{d,syn,max}$ from $T_{d,model}$ for quantitative comparison. Figure 1 presents examples of the averaged normalized deviation of $T_{d,syn,max}$ from $T_{d,model}$ within $20~AU<r<45~AU$, that is, $(1/n_{20-45AU}) \sum_{r=20AU}^{45AU}(T_{d,syn,max}(r)-T_{d,model}(r))/T_{d,model}(r)$, where $n_{20-45AU}$ is the number of grids among $r=20-45~AU$. The bottom-left triangle shows the results for Band [10,x1,y1] sets and the top-right triangle shows those for Band [x2, y2, 3] sets, where x1=[9,8,7,6,5,4], y1=[8,7,6,5,4,3], x2=[10,9,8,7,6,5], and y2=[9,8,7,6,5,4]. According to the calculation, Band [10,6,3] set gives us $T_{d,syn,max}$ deviates $\sim21\%$ from $T_{d,model}$ in average within $20~AU<r<45~AU$. It is the best band set among all the possible 56 band sets. 

\begin{figure}[t]
     \centering
     \includegraphics[width=0.95\linewidth]{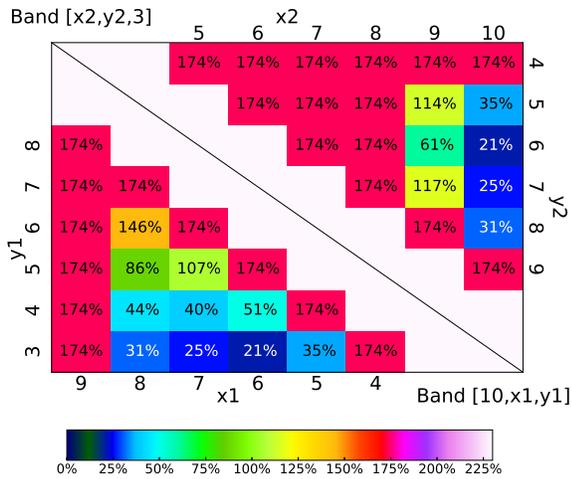}
     \caption{The averaged values of $(T_{d,syn,max}-T_{d,model})/T_{d,model}$ within $20~AU<r<45~AU$. The bottom-left triangle presents the values derived from the Band [10, x1, y1] sets and the top-right triangle presents the values derived from the Band [x2, y2, 3] sets. The background colors of blocks indicate small errors as blue and large errors as red. The constrained $T_{d,syn,max}$ from Band [10,6,3] set is $\sim21\%$ larger than $T_{d,model}$ in average, and this is the best band set.}
     \label{fig:figure1}
\end{figure}

\begin{figure}[t]
     \centering
     \includegraphics[width=\linewidth, clip]{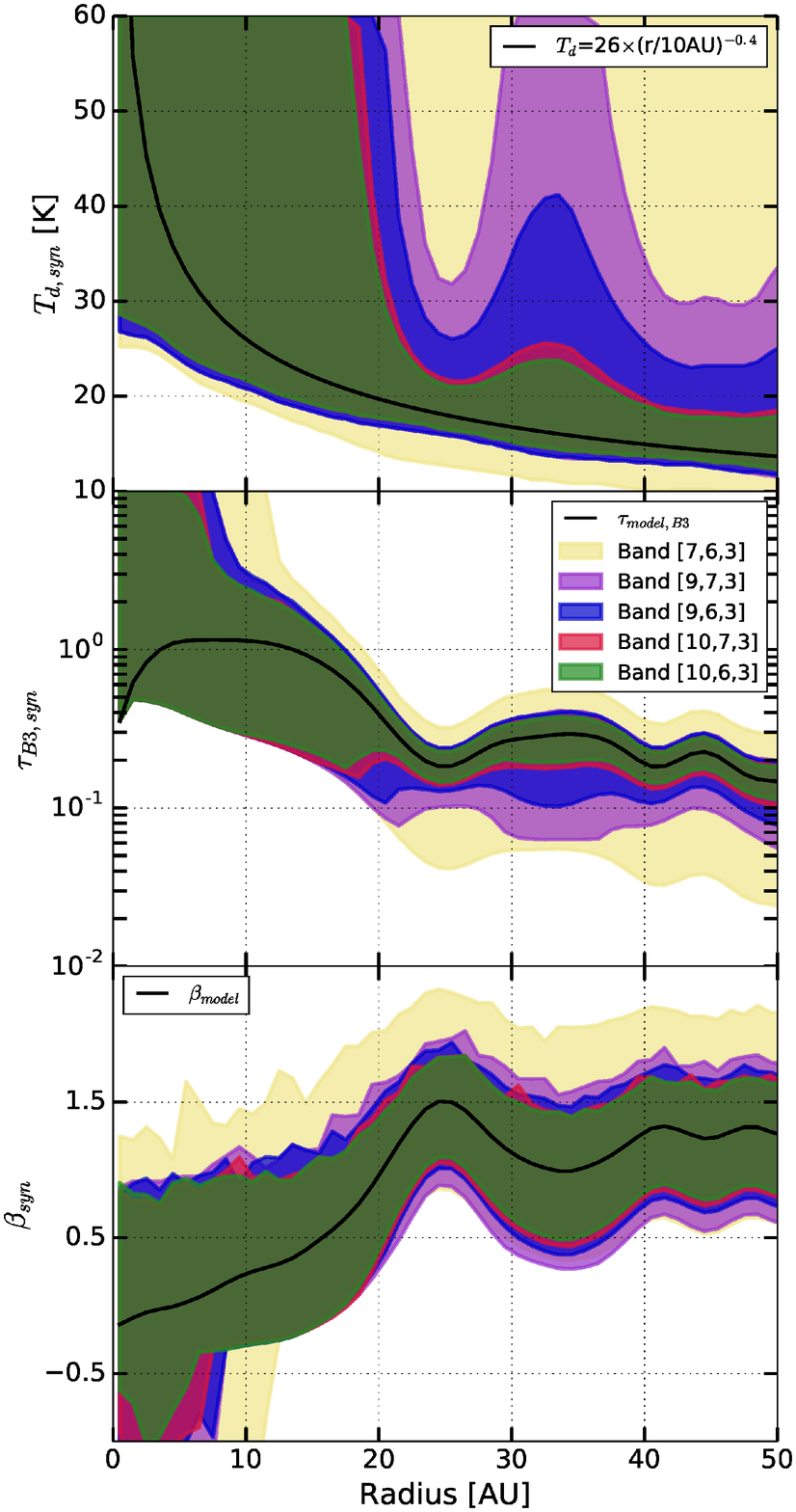}
     \caption{The constraint range of radial profiles of synthetic dust temperature $T_{d,syn}$ (top), synthetic optical depth at ALMA Band 3 $\tau_{B3,syn}$ (middle), and synthetic opacity power-law index $\beta_{syn}$ (bottom) derived from five ALMA band sets using the synthetic multiband analysis. The different colors represent the different ALMA band set; Band [7,6,3] set by yellow, [9,7,3] by purple, [9,6,3] by blue, [10,7,3] by red, and [10,6,3] by green. The Black solid line shows the model radial profile of $T_{d,model}$, $\tau_{B3,model}$, and $\beta_{model}$ derived by the method described in Section \ref{subsec:model}. 
     }%
     \label{fig:figure2}
\end{figure}

\begin{figure*}[t]
     \centering
     \includegraphics[width=0.95\linewidth, clip]{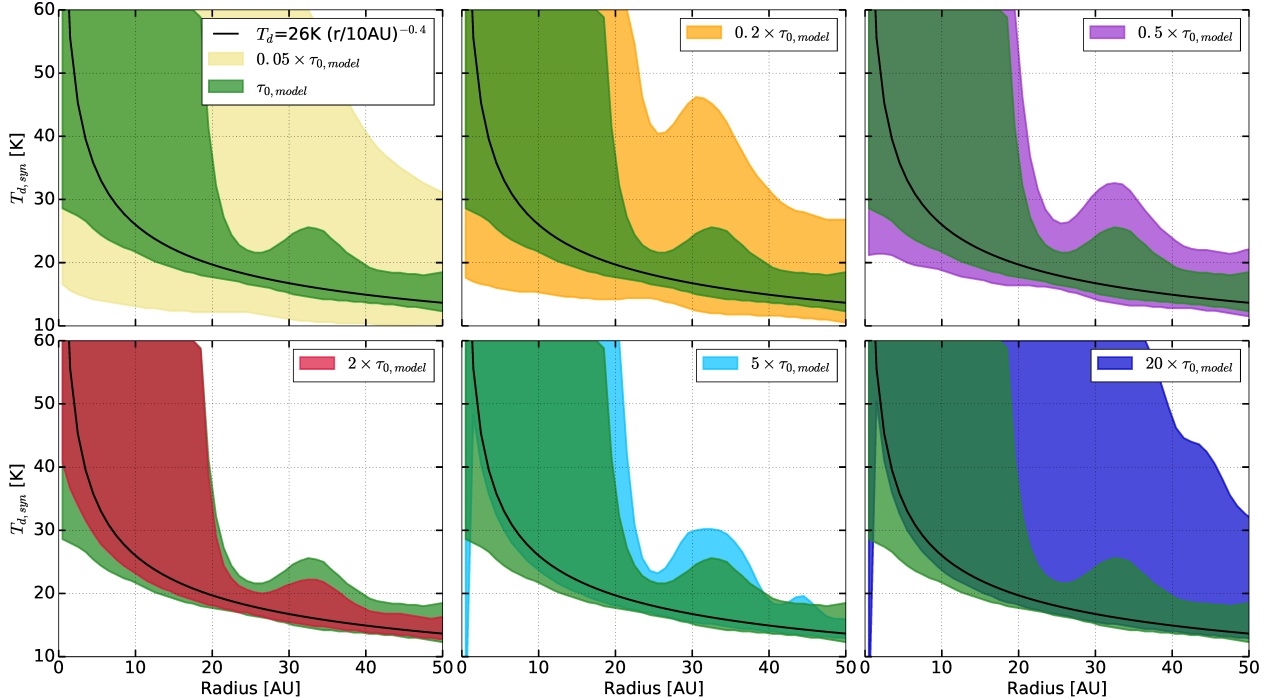}
     \caption{The constraint ranges of $T_{d,syn}$ radial profiles derived from Band [10,7,3] set with the different model profile of optical depth $\tau'_{0,model}$. From Top-left panel, the results derived from $\tau'_{0,model}=0.05\times\tau_{0,model}$ (yellow), $0.2\times\tau_{0,model}$ (orange), $0.5\times\tau_{0,model}$ (purple), $2\times\tau_{0,model}$ (red), $5\times\tau_{0,model}$ (lightblue), and $20\times\tau_{0,model}$ (blue) are presented. Green shaded region indicates the derived $T_{d,syn}$ range with the original $\tau_{0,model}$. As optical depth decreases, the constraint range of $T_{d,syn}$ becomes larger. On the other hand, the slightly enhanced optical depth improves the constraints on synthetic dust temperature while extremely optically thick limits also become worse on synthetic dust temperature constraints. }%
     \label{fig:figure3}
\end{figure*}

Based on the analysis in Figure 1, Figure 2 shows the constraint ranges of the synthetic dust properties of four good ALMA band sets ([10,7,3], [10,6,3], [9,7,3], [9,6,3]) and one bad ALMA band set ([7,6,3]) as examples selected from Figure 1. The constraint ranges of synthetic dust temperature $T_{d,syn}$ (top), synthetic optical depth at Band 3 $\tau_{B3,syn}$ (middle), and synthetic dust opacity power-law index $\beta_{syn}$ (bottom) derived by the synthetic multiband analysis are plotted by different color-shaded regions; yellow indicates Band [7,6,3] set, purple is [9,7,3] set, blue is [9,6,3] set, red is [10,7,3] set, and green is [10,6,3] set. The black line in each panel indicates the assumed model radial profile of $T_{d,model}(r)$ and the derived model radial profile of $\tau_{B3,model}(r)$ and $\beta_{model}(r)$, respectively (see Section \ref{subsec:model}). 

In Figure 2, Band [10,6,3] set is the best set for constraining dust properties. Then, the better results are derived from the following band sets in the order of Band [10,7,3] set, [9,6,3] set, and [9,7,3] set. Including other good band sets in Figure 1, they have two common conditions. 
\begin{enumerate}
\item Band 9 or 10 is included in the band set 
\item Enough frequency intervals between the bands
\end{enumerate} 

For the minor accuracy, Band 10 has the priority for the better constraints on the dust properties than Band 9 if the other two bands in the sets are the same. We can see this tendency in Figure 1 and 2 comparing the band sets of [9,7,3] vs [10,7,3] and [9,6,3] vs [10,6,3]. In addition, even for two similar good ALMA band sets, one band set is slightly better than the other. For instance, the Band [9,6,3] set shows better constraint ranges on the synthetic dust properties than the Band [9,7,3] set due to the relatively large frequency intervals between the bands. 

For the band sets which do not satisfy two conditions, the uncertainties of the synthetic dust properties become large. For instance, the band sets including Band 8 instead of Band 9 or 10 (i.e. Band [9,6,3] set vs [8,6,3] set) have broader constraint ranges of dust properties and the upper limit of dust temperature $T_{d,max}\gtrsim 60~K$ at every radius and give an error of 174\% in Figure 1, even though the observational error at Band 8 ($x=15\%$) is smaller than that of Band 9 ($x=20\%$). Meanwhile, even if Band 9 or 10 is included, in the cases that two bands among the three bands are sequential, the upper limit of the derived dust temperature becomes $T_{d,max} > 60~K$ and gives large errors of 174\% in Figure 1.

\subsection{Interpretations of the synthetic analysis results}
\label{subsec:synthetic_interpretation}

According to the results of synthetic ALMA multiband analysis, we find that the ALMA band sets including Band 9 or 10 and having enough frequency intervals between the bands provide us better constraint ranges of $T_{d,syn}$, $\tau_{\nu,syn}$, and $\beta_{syn}$. To understand why those conditions are necessary for good ALMA band sets, we examine the effect of each dust property to see how the behaviors of the synthetic analysis results vary by artificially changing the dust parameters.

\subsubsection{$\tau_{\nu}$ effect}
\label{subsubsec:tau_effect}

We analyze the behaviors of the synthetic multiband analysis results when the optical depth varies by artificially multiplying some factors to the $\tau_{0,model}(r)$ profile (see Section \ref{subsec:model}). Figure 3 illustrates the constraint range of $T_{d,syn}$ derived from Band [10,7,3] set with the artificially adjusted model optical depth profile $\tau'_{0,model}(r)$. From top-left to bottom-right panel, $\tau'_{0,model}(r)$ = 0.05$\times$, 0.2$\times$, 0.5$\times$, 2$\times$, 5$\times$, and 20$\times\tau_{0,model}(r)$. The green shaded region in all panels indicates the $T_{d,syn}$ constraint range derived from the original $\tau_{0,model}(r)$ as the comparison. The black solid line in all panels indicates the model profile of dust temperature $T_{d,model}(r)=26K(r/10AU)^{-0.4}$ . 

Figure 3 shows that it becomes difficult to constrain the dust temperature when the $\tau_{0,model}$ becomes large or small. The range of $\tau_{0,model}(r) \leq \tau'_{0,model}(r) \leq 2\times \tau_{0,model}(r)$ provides good constraints on $T_{d,syn}$ for Band [10,7,3] set. These behaviors can be understood by Equation (1). In the optically thick ($\tau_{\nu}\gg1$) and thin ($\tau_{\nu}\ll1$) limit, Equation (1) is approximated as 
 \begin{equation}
I_{\nu} \approx B_{\nu}(T_{d})
\end{equation}
and
\begin{equation}
I_{\nu} \approx B_{\nu}(T_{d})\tau_{\nu}
\end{equation}
respectively. If the band set contains only optically thick bands, as in the case of $\tau'_{0,model}(r)=20\times \tau_{0,model}(r)$, we are not able to constrain the optical depth, $\tau_{\nu}$, accurately from the observed intensities because Equation (4) does not depend on $\tau_{\nu}$. Meanwhile, if the band set contains only optically thin bands, as in the case of $\tau'_{0,model}(r)=0.05\times \tau_{0,model}(r)$, it is also difficult to constrain the dust properties because the dependence of the observed intensities on the dust temperature and the optical depth are completely degenerated as is shown in Equation (5). Therefore, we can obtain better constraints on the dust properties only when the band set contains both optically thick and thin bands. That's because optically thick bands constrain the dust temperature well as is indicated in Equation (4) and optically thin band constrains the dust optical depth as is indicated in Equation (5).
 
\begin{figure}[t]
     \centering
     \includegraphics[width=\linewidth, clip]{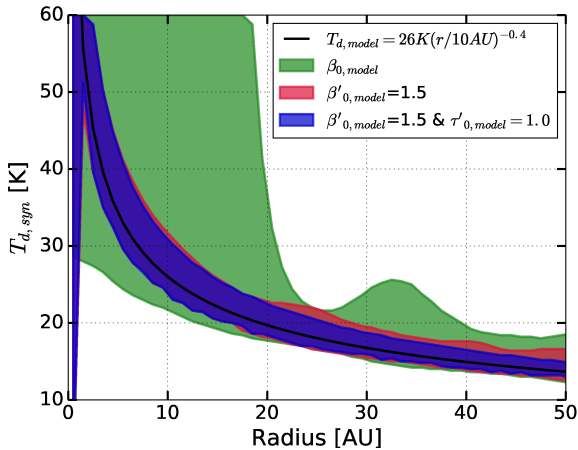}
     \caption{The constraint range of $T_{d,syn}$ profiles derived from Band [10,7,3] set with $\beta'_{model}(r)=1.5$ at every radius (red) and $\beta'_{model}(r)=1.5$ and $\tau'_{0,model}(r)=1.0$ at every radius (blue). The result derived from the original $\beta_{model}$ and $\tau_{0,model}$ profiles is presented by green shaded region. By this setting, $\beta'_{model}(r)\geq\beta_{model}(r)$ at every radius. As $\beta$ increases, we obtain better $T_{d,syn}$ constraint ranges. The blue shaded region shows purely $T_{d}$ effect since we controlled two parameters, $\beta'_{model}$ and $\tau'_{0,model}$.}%
     \label{fig:figure4}
\end{figure}

\begin{figure}[t]
     \centering
     \includegraphics[width=\linewidth, clip]{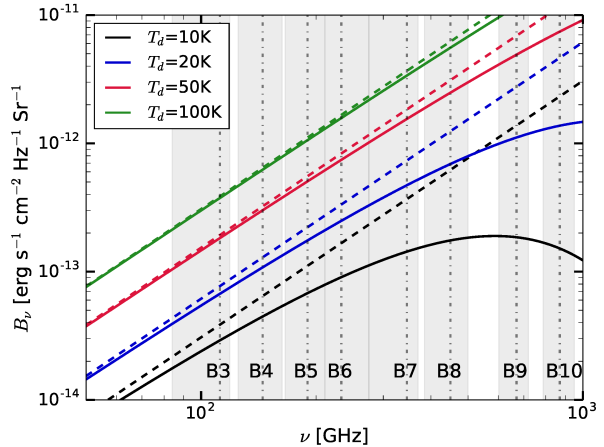}
     \caption{The blackbody curves at the dust temperature of $10~K$ (black), $20~K$ (blue), $50~K$ (red), and $100~K$ (green) on the $log(I_{\nu})$-$log(\nu)$ plot. The dashed lines indicate the formula of the Rayleigh-Jeans limit which satisfies the relation $I_{\nu}\propto \nu^2$. The grey shaded regions present the frequency coverage and the dashed-dotted vertical lines indicate the representative frequency of each ALMA band listed in Table 1. The blackbody curve deviates from the Rayleigh-Jeans limit especially at high frequency (Band 9 and 10) and low temperature ($T_{d}=10-20~K$).}%
     \label{fig:figure5}
\end{figure}

\subsubsection{$\beta$ effect}
\label{subsubsec:beta_effect}

Similarly, we examine the behavior of the synthetic analysis results with artificially adjusted $\beta'_{model}$ profile. We derive the constraint range of the $T_{d,syn}$ with $\beta'_{model}(r)=1.5$ which is $\beta'_{model}(r)\geq\beta_{model}(r)$ at all radii where $\beta_{model}(r)$ is the original model profile. Figure 4 illustrates the constraint range of $T_{d,syn}$ derived from Band [10,7,3] set with the $\beta'_{model}(r)=1.5$ (red) and with the original $\beta_{model}(r)$ (green). The constraint range of $T_{d,syn}$ profile becomes much better as $\beta$ increases even in the inner disk, $r<25~AU$, and at the $T_{d,syn}$ bump around $r\sim33~AU$. 

We can interpret this behavior by the similar way described in Section \ref{subsubsec:tau_effect}. If $\beta$ is large, the optical depth becomes thick at high-frequency bands and thin at low-frequency bands. Therefore, we can obtain good constraints on the dust properties from the observed intensities using Equations (1) and (2). Meanwhile, if $\beta$ is small, the optical depths become similar at all the bands. That is, all the bands become optically thick or thin, and then we are not able to constrain the dust properties accurately.

\subsubsection{$T_{d}$ effect}
\label{subsubsec:Td_effect}

In Figure 4, the blue shaded region shows the constraint range of $T_{d,syn}$ derived with $\tau'_{0,model}(r)=1.0$ and $\beta'_{model}(r)=1.5$ at every radius, which reflects only the effect of the dust temperature. Meanwhile, The red shaded region contains the dependence on the dust opacity as well, for example, the errors become slightly large at the gap regions where the optical depths become small. This result shows that the error becomes larger if the dust temperature is higher. It is explained by the deviation of blackbody curves from the Rayleigh-Jeans (RJ) limit. 

Figure 5 presents the blackbody curves at the dust temperature of $10~K$ (black), $20~K$ (blue), $50~K$ (red), and $100~K$ (green) represented by solid lines, respectively. The dashed linear lines indicate the formula of the RJ limit where $I_{\nu}\propto \nu^2$. The grey shaded regions indicate the frequency coverage and the vertical dashed-dotted lines are the representative frequencies of each band listed in Table 1. For the very cold dust temperature ($T_{d}\leq20~K$), the peak of the blackbody curves locates at the frequency regimes of ALMA Band 9 or 10. Thus, the blackbody curves largely deviate from the RJ limit at Band 9 and 10. If the deviation from the RJ limit becomes larger, the dust temperature can be constrained well from the slope of the observed intensities. This is one of the reasons why we need to include Band 9 or 10 in order to obtain better constraints on the dust properties. In addition, since the deviation from the RJ limit at Band 10 is larger than one at Band 9, Band 10 data gives more precise constraints than Band 9 data.

\begin{figure*}[p]
     \centering
     \includegraphics[width=\linewidth, clip]{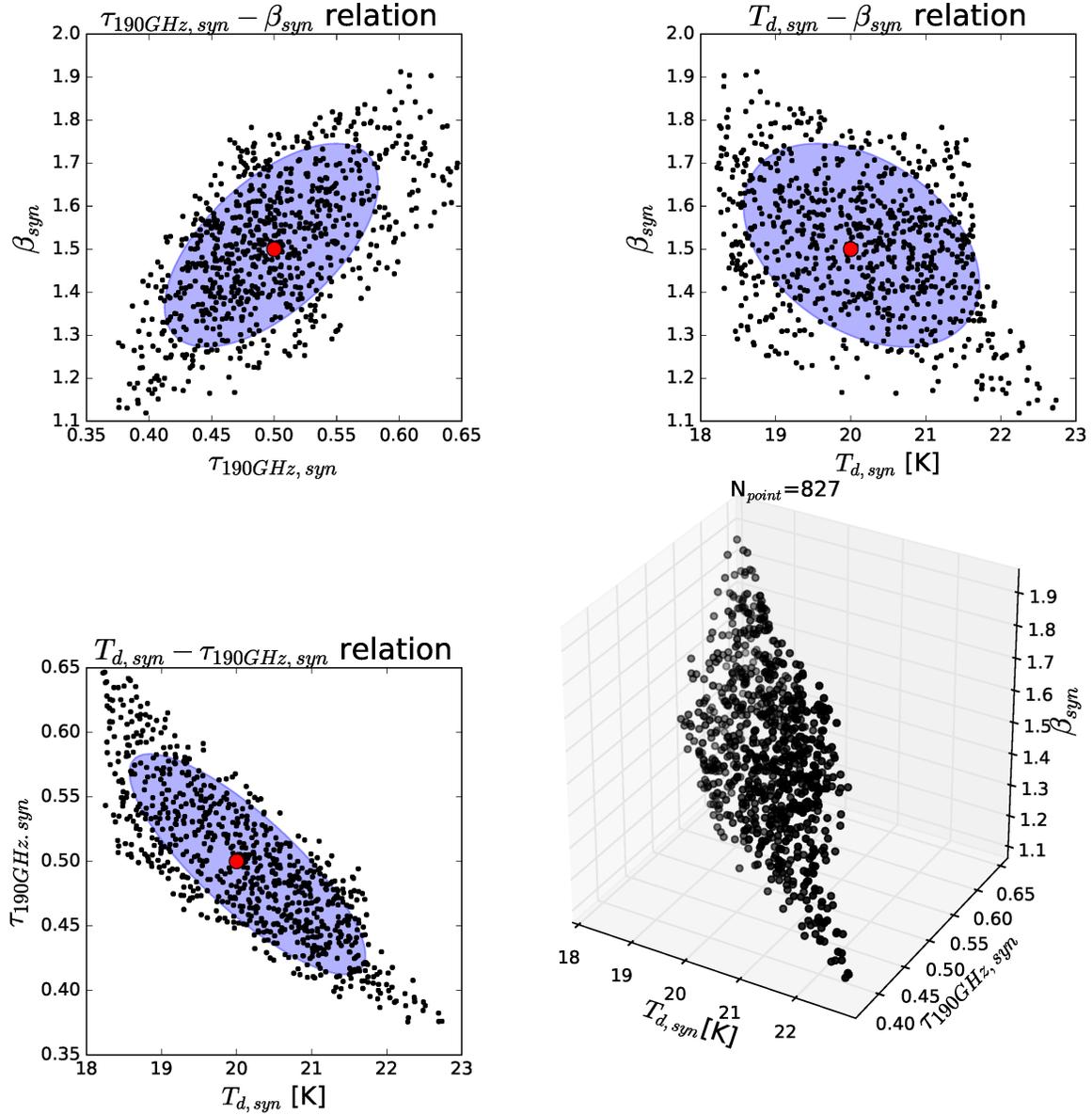}
     \caption{The 3D scatter plot of [$T_{d,syn}$, $\tau_{190GHz,syn}$, $\beta_{syn}$] derived from randomly generated intensities at ALMA Band [10,6,3] set (bottom-right) and their projected 2D scatter plots. The generated intensities have random errors of $\pm10\%$ for Band 6 and 3 and $\pm20\%$ for Band 10 against the model intensities derived from [$T_{d,model}=20~K$, $\tau_{190GHz,model}=0.5$, $\beta_{model}=1.5$]. The red points in 2D scatter plots are the model values. The blue ellipse in 2D scatter plots is the $1\sigma$ confidence ellipse fitting which contains 68\% of all points. The points show correlations between dust properties: $\tau_{190GHz,syn}$ and $\beta_{syn}$ are getting smaller as  $T_{d,syn}$ becomes larger.}%
     \label{fig:figure6}
\end{figure*}

\begin{figure*}[p]
\centering  
\subfiguretopcaptrue
\mbox{}
\includegraphics[width=0.8\linewidth, clip]{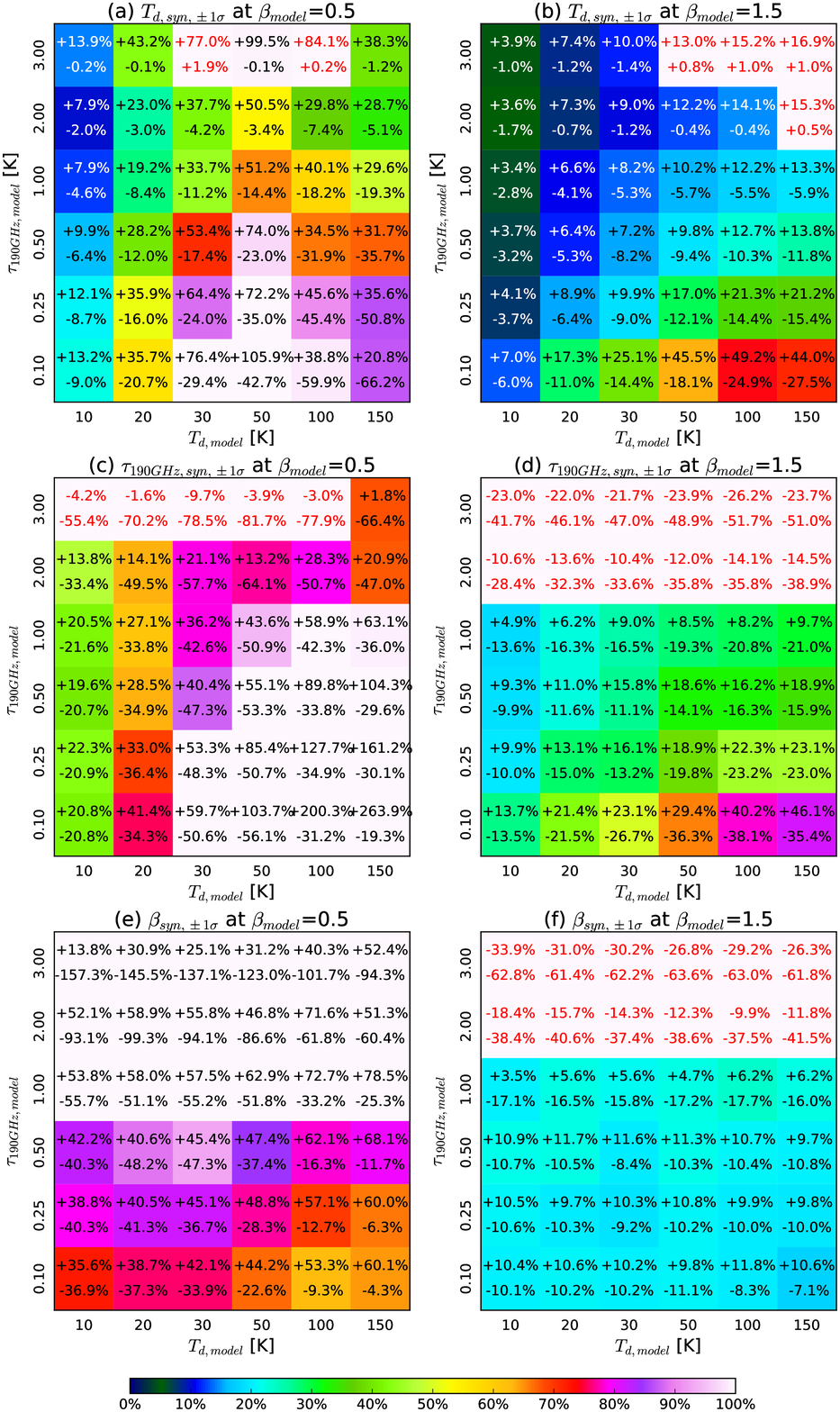}
\caption{Errors in the derived $T_{d,syn}$ (a and b), $\tau_{190GHz,syn}$ (c and d), and $\beta_{syn}$ (e and f) for various model values of [$T_{d,model}$, $\tau_{190GHz,model}$, $\beta_{model}$] at Band [10,6,3] set. The percentages written in each block indicate the normalized deviation of $+1\sigma$ (upper) and $-1\sigma$ (lower) values of the derived dust properties from the model values. The red numbers indicate the cases that the model value is not included in the $\pm1\sigma$ range. Background colors indicate the deviation between $\pm 1\sigma$ values normalized by the model value. The figure shows that we obtain better constraints on dust properties with low $T_{d,model}$, intermediate $\tau_{190GHz,model}$, and high $\beta_{model}$. This is consistent with the interpretations described in Section \ref{subsec:synthetic_interpretation}.}
\label{fig:figure7}
\end{figure*}

\subsubsection{Frequency intervals between Bands}

The frequency intervals between the bands of the set also affect the accuracy of the derived dust properties. Since spectral index $\alpha$ is defined as $\alpha\equiv\Delta log(I_{\nu})/\Delta log(\nu)$, the error of $\alpha$ becomes smaller when the frequency interval $\Delta log(\nu)$ increases for a given $\Delta log(I_{\nu})$. As shown in Equation (3) in Section \ref{subsec:model}, $\alpha$ is strongly related with the dust properties, $T_{d}$, $\tau_{\nu}$ and $\beta$, and thus we can derive the dust properties from the observed intensities. If we adopt $B_{\nu}(T)\propto \nu^{\gamma}$ at a certain frequency and a certain temperature, $\alpha \approx \gamma$ in the optically thick limit, which corresponds to Equation (4). In the optically thin limit, $\alpha \approx \gamma+\beta$, which corresponds to Equation (5). We note that $\gamma= 2$ in the RJ limit. Therefore, the error of $\alpha$ directly affects the constraints on the dust properties. Due to this effect, we need large frequency intervals between Bands for the good ALMA band sets to reduce the error of $\alpha$.

In principle, this interpretation suggests that we can extend this analysis to the longer or shorter wavelengths beyond the ALMA to obtain larger frequency intervals between the data. However, we should be cautious about the differences in spatial resolution and sensitivity when we combine the data observed by different telescopes. For example, VLA is another good instrument to obtain longer wavelength data but the spatial resolution and sensitivity may not be comparable to ALMA data.

\subsection{The generalization of the synthetic multiband analysis}
\label{subsec:general_conditions}

We investigate the physical reasons of two conditions for the good ALMA band sets in Section \ref{subsec:synthetic_interpretation}. Here, we apply the synthetic multiband analysis to more generalized cases adding Monte-Carlo (MC) method to mimic the observations, not restricted on the TW Hya disk. To do this analysis, we focus on the ALMA Band [10,6,3] set for minimizing the loss of the generated random data because it is the best one in the synthetic multiband analysis.

\subsubsection{The degeneracy of [$T_{d}$, $\tau_{\nu}$, $\beta$]}
\label{subsubsec:degeneracy}

The interpretations investigated in Section \ref{subsec:synthetic_interpretation} infer that the dust properties have degeneracy on the multiband analysis. First, to examine their degeneracy on the synthetic multiband analysis, we survey the parameter space by Monte-Carlo method for a certain combination of the model dust properties, [$T_{d,model}$, $\tau_{190GHz,model}$, $\beta_{model}$]. To do this parameter survey, we calculate the model intensities at ALMA Band 10, 6 and 3 corresponding to the model dust properties using Equation (1) and (2) in Section \ref{subsec:model}. Then, we generate 1000 combinations of the synthetic intensity with the random errors within $\pm10\%$ for Band 6 and 3 and $\pm20\%$ for Band 10. And we derive the dust properties using the multiband analysis. 

Figure 6 presents 3-dimensional (3D) scatter plot of the derived [$T_{d,syn}$, $\tau_{190GHz, syn}$, $\beta_{syn}$] and 2-dimensional (2D) scatter plots projected to each axis for examining their degeneracy. The 827 points are plotted from 1000 combinations of the randomly generated synthetic intensities for [$T_{d,model}$,$\tau_{190GHz,model}$,$\beta_{model}$]=[20 $K$, 0.5, 1.5]. Since some intensity combinations don't have the solutions, we lose some portion of them. The red points in 2D scatter plots indicate this model value. With these conditions, the black points are concentrated in a diagonal line in the 3D scatter plot of Figure 6. The 2D scatter plots of Figure 6 show the projected correlation between dust properties. The blue ellipses in 2D scatter plots indicate the $1\sigma$ confidence ellipse which contains $\sim68\%$ of the plotted points. They indicate the correlation direction and their strength. Based on it, we confirm $T_{d,syn}$ has negative correlations to $\tau_{190GHz,syn}$ and $\beta_{syn}$ and $T_{d,syn}-\tau_{190GHz,syn}$ has the strongest correlation. We can understand this trend as follows. By Equation (1) in Section \ref{subsec:model}, $\tau_{190GHz}$ should be smaller to cancel out the increment of $B_{\nu}(T_{d})$ as $T_{d}$ increases for given intensities. At the same time, when $T_{d}$ becomes higher, the deviation from RJ limit gets smaller (see Section \ref{subsubsec:Td_effect}). To match the observed spectral index of intensities at three ALMA bands, $\beta$ should be smaller to cancel out the effect of $T_{d}$ variations. Thus, $\tau_{190GHz}$ and $\beta$ have a positive correlation.

\subsubsection{The synthetic analysis for general disk conditions}
\label{subsubsec:generalize}

Next, by changing the model values, we examine the applicability of the multiband analysis to various situations. We set some representative model dust properties as $T_{d,model}$=[10$K$, 20$K$, 30$K$, 50$K$, 100$K$, 150$K$], $\tau_{190GHz,model}$=[0.1, 0.25, 0.5, 1.0, 2.0, 3.0], and $\beta_{model}$=[0.5, 1.5]. Those parameters cover a wide range of possible disk conditions, from cold ($10~K$) to hot ($150~K$), from optically thin ($\tau_{190GHz}\leq0.5$) to thick ($\tau_{190GHz}\geq1.0$), and with small grains ($\beta=1.5$, $a_{max}\lesssim1~mm$) and large grains ($\beta=0.5$, $a_{max}\gtrsim1~cm$) \citep[e.g.][]{Draine2006}. 

We perform the same calculation applied to Figure 6 for various combinations of the model values. Then, we make the occurrence histogram of the derived $T_{d,syn}$, $\tau_{190GHz,syn}$, and $\beta_{syn}$, and then pick their $\pm 1 \sigma$ values (the values at $\pm 34\%$ from the mean values of all the derived points of dust properties), $T_{d,syn,\pm 1\sigma}$, for example. Figure 7 summarizes the normalized $\pm 1\sigma$ values of the derived dust properties from the model values, that is, $(T_{d,syn,\pm 1\sigma}-T_{d,model})/T_{d,model}$, for example. The normalized $T_{d,syn,\pm1\sigma}$ (top row), $\tau_{190GHz,syn,\pm1\sigma}$ (middle row), and $\beta_{syn,\pm1\sigma}$ (bottom row) are shown for a certain set of [$T_{d,model}$,$\tau_{ 190GHz,model}$,$\beta_{model}$] in each block where x-axis shows $T_{d,model}$ and y-axis shows $\tau_{ 190GHz,model}$. $\beta_{model}=0.5$ and 1.5 are used for the figures in the left and right columns, respectively. The numbers in the upper and lower parts in each block show the normalized deviation of the $+1\sigma$ and $-1\sigma$ values, respectively.
We note that the red numbers indicate the cases in which the model value is not included in $\pm1\sigma$ range, that is, the model value is not well reproduced by the multiband analysis. It occurs when large errors exist in the derived dust properties, for example, in the case of large $\tau_{190GHz,model}$. The background colors indicate the total width between $\pm 1\sigma$ values normalized by the model value, that is, $(A_{syn,+1\sigma}-A_{syn,-1\sigma})/A_{model}$ where $A$ is $T_{d}$, $\tau_{190GHz}$, or $\beta$. For example, in Figure 7 (a), for the model values are [$T_{d,model}$,$\tau_{190GHz,model}$,$\beta_{model}$]=[20 $K$, 1.0, 1.5], $T_{d,syn,+1\sigma}\approx21.32~K$ ($+6.6\%$) and $T_{d,syn,-1\sigma}\approx19.18~K$ ($-4.1\%$). Therefore, the normalized total width is $\sim$10.7\%.

 According to the results, we obtain good constraints on dust properties with low $T_{d,model}$, intermediate $\tau_{190GHz,model}$, and high $\beta_{model}$. This is consistent with the interpretations in Section \ref{subsec:synthetic_interpretation}. In addition, $\tau_{190GHz,model}$ and $\beta_{model}$ have more strict ranges for good constraints than $T_{d,model}$. That is, making the combination of optically thick and thin bands is the primary factor for good band sets. For $\tau_{190GHz,model}\geq2.0$, it is a situation that Band 10, 6 and 3 are all optically thick which is not suitable for the multiband analysis. Due to the degeneracy on $T_{d}$ and $\tau_{\nu}$ in Equation (1), the derived constraints on dust properties are not able to reproduce the model values. 

\begin{figure*}[t]
\centering  
\subfiguretopcaptrue
\mbox{}\hfill  
\subfigure[The azimuthally averaged radial intensity profiles]{\label{fig:a}\includegraphics[width=0.45\textwidth]{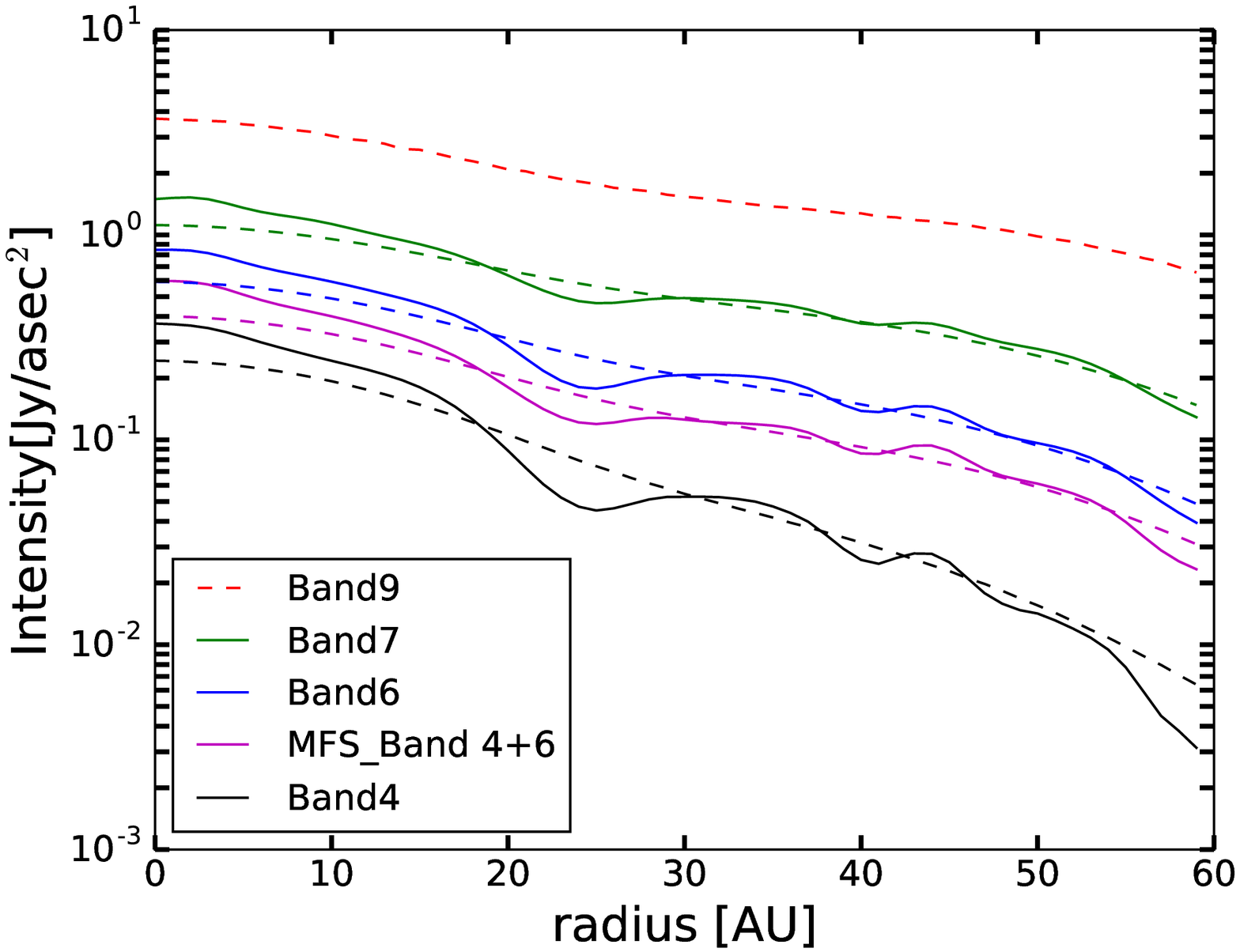}}\hfill
\subfigure[The brightness temperature radial profiles]{\label{fig:b}\includegraphics[width=0.45\textwidth]{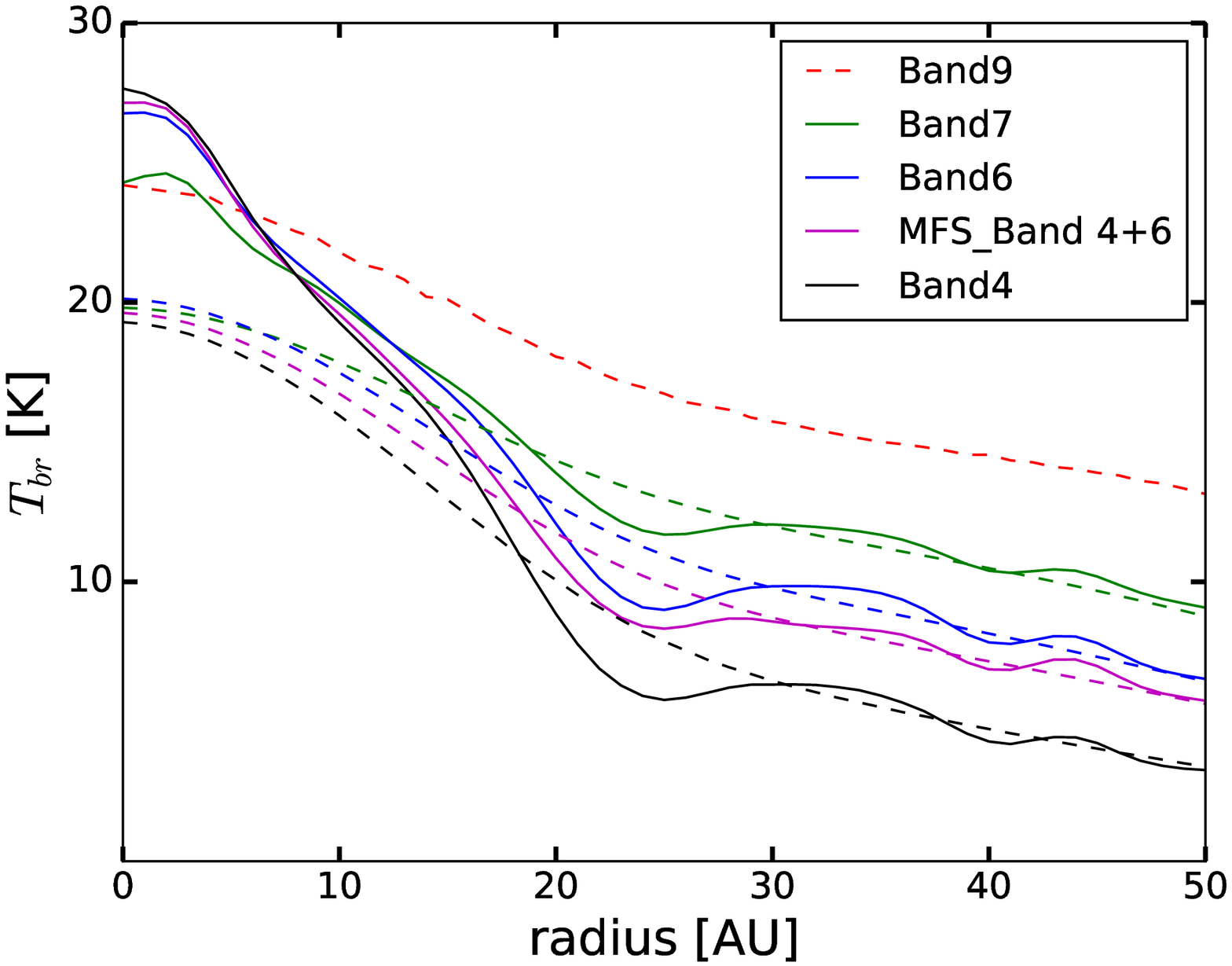}}
\hfill\mbox{}
\subfiguretopcapfalse
\caption{{\bf (a)} The azimuthally-averaged de-projected radial intensity profiles and {\bf (b)} the brightness temperature derived from these intensity profiles. The different color indicates different ALMA band; Band 4 (black), 6 (blue), 7 (green), and 9 (red). The magenta indicates the profiles obtained by the multi-frequency synthesis (MFS) method from Band 4 and 6. The solid lines are the profiles of high-resolution data ($\sim88.1\times62.1~mas$) and the dashed lines are the profiles of low-resolution data ($\sim434\times247~mas$).}
\label{fig:figure8}
\end{figure*}

\begin{figure*}[t]
     \centering
     \includegraphics[width=\linewidth, clip]{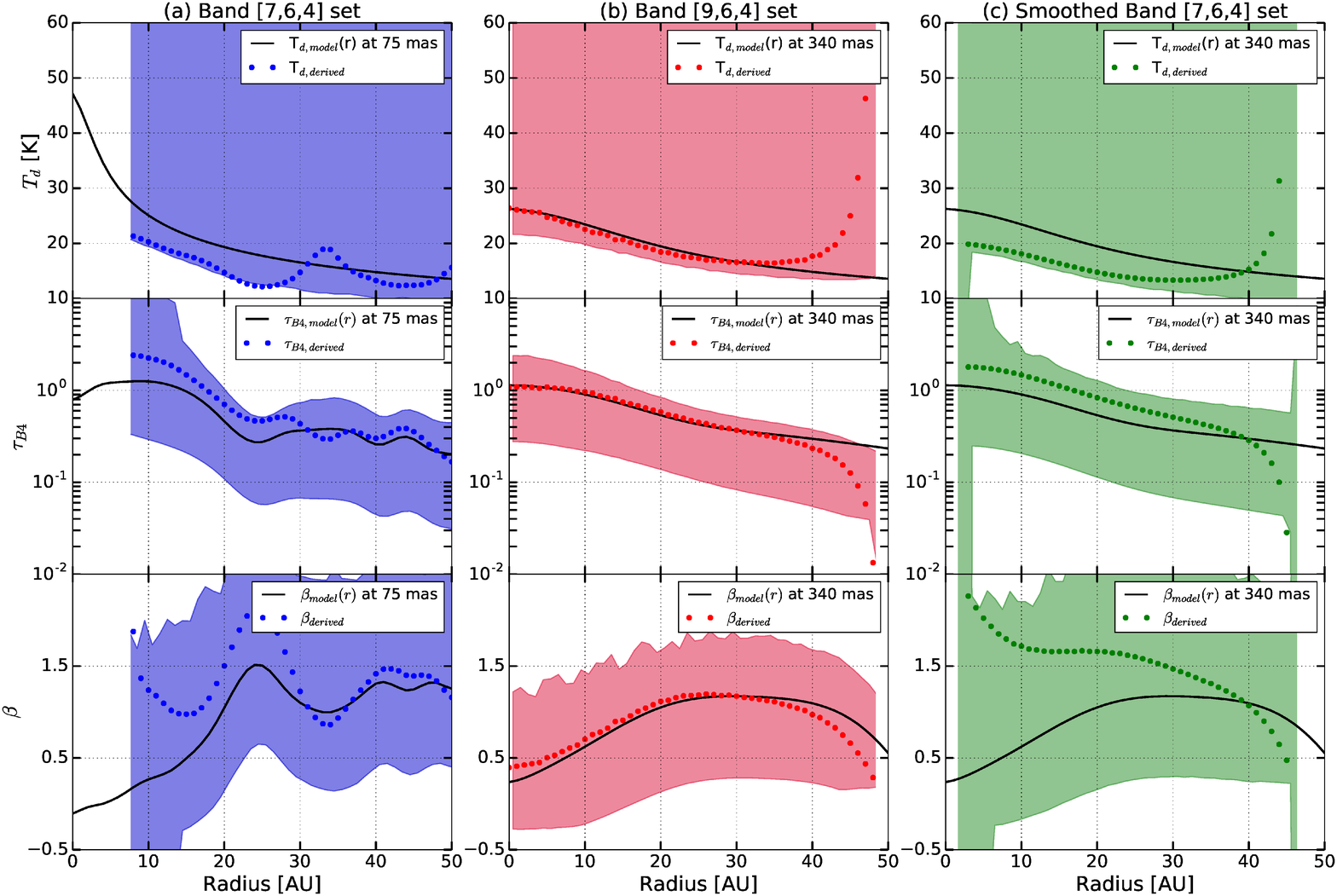}
     \caption{The dotted lines show $T_{d}$ (top), $\tau_{B4}$ (middle), and $\beta$ (bottom) radial profiles derived from Band [7,6,4] set (left), Band [9,6,4] set (center), and the smoothed Band [7,6,4] set (right) by the ALMA multiband analysis. The black solid lines are the model profiles of  $T_{d,model}$, $\tau_{B4,model}$, and $\beta_{model}$ at $\sim75$ mas (left) and $\sim340$ mas (center and right) circular beam. $T_{d,model}(r)$ is derived from $T_{d}(r)= 26K(r/10AU)^{-0.4}$ by convolution. The frequency independent $\tau_{B4,model}(r)$ and $\beta_{0,model}(r)$ profiles are derived using $T_{d,model}(r)$ (see Section \ref{subsec:model}). The color-shaded regions indicate the constraint ranges of the dust properties corresponding to the $\pm10\%$ observational error for Band 4 and 6, $\pm15\%$ for Band 7, and $\pm20\%$ for Band 9.}%
     \label{fig:figure9}
\end{figure*}

\section{Applications of the multiband analysis to ALMA archival data}
\label{sec:application}

In Section \ref{sec:results}, we present the synthetic multiband analysis results and find two conditions for the good ALMA band sets deriving accurate constraints on the dust properties. In this section, we apply the multiband analysis to ALMA archival dust continuum data of the TW Hya protoplanetary disk at Band 4, 6, 7, and 9 in order to derive the dust properties and to examine whether our synthetic multiband analysis results are consistent with the real observational data. 

\citet{Tsukagoshi2016} and \citet{Andrews2016} reported the high spatial resolution dust continuum images of the TW Hya disk at Band 4, 6 and 7. We use Band 4 and 6 data in \citet{Tsukagoshi2016} and Band 7 fits image at Sean Andrews's science homepage\footnote{https://www.cfa.harvard.edu/~sandrews/data/twhya/}. In the ALMA archive, we also find Band 9 observation of the TW Hya disk \citep{Schwarz2016} which is low spatial resolution data. We retrieved this data from JVO portal\footnote{http://jvo.nao.ac.jp/portal/alma/archive.do} operated by the NAOJ. We summarize the beam size and 1$\sigma$ noise level of those archival data in Table 2. We make two ALMA band sets, Band [7,6,4] set as high-resolution data and Band [9,6,4] set as low-resolution data. 

\begin{table}[b]					
\begin{center}
\label{Table2}											
\caption{The basic information of ALMA archival data used for the multiband analysis}
\begin{tabular*}{0.45\textwidth}{@{\extracolsep{\fill}} c | c c c c }			
\hline \hline					
\multirow{2}{*}{Band}	  &	$\nu_{rest}$ & Beam size	 &	PA & 1$\sigma$ level	\\
 	&  [GHz] 	&	[mas$\times$mas]	& 	[$^{\circ}$]	&	[mJy/beam] 	\\
\hline					
4	&	145 & $\sim$88.1$\times$62.1 & $\sim$57.8 & $\sim$ 0.0124	\\
\hline
6	&	233 & $\sim$75.4$\times$55.2 & $\sim$38.0 & $\sim$ 0.0287 \\
\hline
7	&	346 & $\sim$24$\times$18 & $\sim$78 & $\sim$ 0.0350 \\
\hline
9	& 	661 & $\sim$434$\times$247 & $\sim$-85.8 & $\sim$ 10 \\
\hline \hline					
\end{tabular*}	
\end{center}					
\end{table}

For the consistent analysis, we match the beam size of the continuum images in each set using $imsmooth$ command in CASA. We match the beam size of $\sim88.1\times62.1$ mas for Band [7,6,4] set and $\sim434\times247$ mas for Band [9,6,4] set. After matching the beam size, we extract the azimuthally-averaged de-projected radial intensity profiles. Figure 8 (a) presents the azimuthally-averaged de-projected radial intensity profiles by the different colors; black for Band 4, blue for Band 6, green for Band 7, red for Band 9, and magenta for the intensity profile at 190 $GHz$ obtained by the Multi-Frequency Synthesis (MFS) method from Band 4 and 6 \citep{Tsukagoshi2016}. The solid lines indicate the radial profiles of high-resolution ($\sim88.1\times62.1$ mas) data and the dashed lines indicate the profiles of low-resolution ($\sim434\times247$ mas) data. Figure 8 (b) presents the brightness temperatures derived from the radial intensity profiles in Figure 8 (a).


\subsection{Application to Band [7,6,4] set}
\label{subsubsec:764set}

We apply the multiband analysis to the azimuthally-averaged de-projected radial intensity profiles of Band [7,6,4] set. The dust properties are derived from the intensity at Band 7 (346 $GHz$) and the intensity and spectral index $\alpha$ at 190 $GHz$ obtained by the MFS method from Band 4 and 6 data. The radial profiles of $T_{d}$, $\tau_{B4}$, and $\beta$ derived from Band [7,6,4] set are shown in Figure 9 (a) as blue dotted lines. These dotted lines are used as the model profiles to derive the blue shaded region which indicates the constraint range of the derived dust properties corresponding to the $\pm10\%$ observational errors for Band 4 and 6 and $\pm15\%$ for Band 7. The black solid line in each panel indicates the model profiles of the dust properties. The $T_{d,model}(r)$ profile is derived from the profile $T_{d}(r)=26K~(r/10~AU)^{-0.4}$ by convolution to $\sim75$ mas circular beam, which is the mean value of $\sim88.1\times62.1$ mas. The frequency independent model profiles $\tau_{B4,model}(r)$ and $\beta_{model}(r)$ are derived using the convolved $T_{d,model}(r)$ by the same method mentioned in Section \ref{subsec:model}. 

In the top panel, the dotted lines are lower than the model profile, $T_{d}(r)\leq20~K$ at $r\gtrsim10~AU$, with a strange bump around $r\sim33~AU$. The bump seems to appear because the uncertainty becomes large around $r\sim33~AU$ due to small $\beta$ as is shown in Section \ref{subsubsec:beta_effect}. The derived optical depth at Band 4 radial profile, $\tau_{B4}(r)$, is close to the model profile, especially at $r>35~AU$, including the shallow drop at the gap locations, $r\sim25~AU$ and $40~AU$. The derived $\tau_{B4}(r)$ values at $r>20~AU$ are in $0.1\lesssim\tau_{B4}(r)\lesssim1$ range and optically thin. The derived $\beta(r)$ profile is also similar to the model profile, especially at $r>30~AU$, but the enhancement at the inner gap location ($r\approx25~AU$) is $\approx40\%$ larger than the model. At $r \leq 15~AU$ where $\tau_{B4}(r) \geq 2$, the deviation becomes very large. These shifts of the dust properties are consistent with the degeneracy described in Section \ref{subsubsec:degeneracy}.

In the inner disk, $r\leq10~AU$, the dust properties are not able to be estimated by the multiband analysis. In this region, the brightness temperatures of Band 4 and 6 are higher than that of Band 7 (see Figure 8 (b)). When the images are convolved with the beam $\gtrsim200$ mas, this inversion of the brightness temperature disappears. One possible cause of it is a significant phase noise of Band 7. Since phase noise makes a positional offset, the resultant image has been blurred and the peak intensity becomes smaller. 

According to the interpretation of the synthetic multiband analysis, Band [7,6,4] set is not a good band set because it does not satisfy two conditions for the good ALMA band sets. The $T_{d}$, $\tau_{B4}$ and $\beta$ profiles derived from Band [7,6,4] set seem to be affected by the errors in the observations, which is consistent with our synthetic analysis results in Section \ref{sec:results}. Also, the constraint range of $T_{d}(r)$ derived from Band [7,6,4] set is too broad to specify the dust temperature at every radius. The constraint ranges of $\tau_{B4}(r)$ and $\beta(r)$ profiles also have relatively large uncertainties at all radius. 

\subsection{Application to Band [9,6,4] set}
\label{subsubsec:964set}

We apply the same method to the low-resolution dust continuum data of the TW Hya disk, Band [9,6,4] set. The dust properties are derived from the intensity at Band 9 (661 $GHz$) and the beam-convolved intensity and spectral index $\alpha$ at 190 $GHz$ obtained by MFS method from Band 4 and 6 data. We derive the radial profiles of $T_{d}(r)$, $\tau_{B4}(r)$, and $\beta(r)$ from those smoothed radial intensity profiles presented by the dashed lines in Figure 8 (a). 

The derived $T_{d}$, $\tau_{B4}$, and $\beta$ radial profiles are shown in Figure 9 (b) as red dotted lines. The format of the plots are the same as Figure 9 (a). The $T_{d,model}(r)$ profile is derived from $T_{d}(r)=26K~(r/10~AU)^{-0.4}$ convolved with $\sim340$ mas circular beam, which is the mean value of $\sim434\times247$ mas. The $\tau_{B4,model}(r)$ and $\beta_{model}(r)$ profiles are derived using this convolved $T_{d,model}(r)$ profile by the same method described in Section \ref{subsec:model}. 

The derived $T_{d}(r)$ profile from Band [9,6,4] set follows the model profile well having $T_{d}(r)\approx15-27~K$ in $r\leq40~AU$ region. Similarly, the derived $\tau_{B4}(r)$ and $\beta(r)$ profiles are also very close to the model $\tau_{B4,model}(r)$ and $\beta_{model}(r)$ profiles. Meanwhile, the derived dust properties deviate from the model in the outer disk, $r>40~AU$, probably due to the rapid drop of $\beta$ (see Section \ref{subsubsec:beta_effect}). The rapid drop of $\beta$ seems to be affected by the drop of $\alpha$ at 190 $GHz$, which is obtained by the MFS method, at $r>50~AU$ where the Signal-to-Noise ratios of the observed intensities become small. By the beam convolution, this effect can be extended to the $r > 40~AU$ region.
 
We derived red shaded regions based on the red dotted lines as the model profiles of the dust properties. The red shaded regions in each panel indicate the constraint ranges of the derived dust properties corresponding to the $\pm10\%$ observational errors for Band 4 and 6 and $\pm20\%$ for Band 9. While the constraint range of $T_{d}(r)$ is still too broad at all radii, it has relatively narrow constraint range of $\tau_{B4}(r)$ and $\beta(r)$ profiles than the ranges derived from Band [7,6,4] set. It is because that Band [9,6,4] set satisfies two conditions for the good band sets; It contains Band 9 in the band set and the frequency intervals between the bands are large. Thus, we confirm that the results derived from the archival Band [9,6,4] set seems less affected by the observational errors than those derived from Band [7,6,4] set, which is consistent with our synthetic analysis results in Section \ref{sec:results}.

The radial profiles of the derived dust properties are smooth because the $T_{d}$ bump at $r\sim33~AU$ and the drop of $\tau_{B4}$ and enhancement of $\beta$ at the gap locations ($r\approx25~AU$ and $\approx40~AU$) are smoothed out. It is because of the large difference in the beam sizes between Band 9 and the others. To check whether those small-scale structures are smoothed out by the large beam size, we derive the dust properties from the smoothed radial intensity profiles at Band [7,6,4] set at $\sim434\times247~mas$ beam size. Figure 9 (c) presents the radial profiles of the dust properties derived from the smoothed radial intensity profiles at Band [7,6,4] set as green dotted lines. The derived $T_{d}(r)$ profile has no bump around $r\sim33~AU$. The $\tau_{B4}(r)$ and $\beta(r)$ profiles also have no structure at the gap location ($r\approx25~AU$). It infers that the small-scale structures in the disk are smoothed out by the large beam size. Therefore, we need a high spatial resolution observation at Band 9 for constraining the dust properties in the gap structures of the TW Hya disk. 

\section{Discussion}
\label{sec:discussion}

In Section \ref{sec:application}, we describe the dust properties derived from the ALMA archival dust continuum images at Band 4, 6, 7, and 9. The results derived from the archival data are well consistent with the synthetic multiband analysis. In this section, we will discuss some points to improve our multiband analysis in the future.

\subsection{The vertical structure of the disk}

In the processes of the synthetic multiband analysis, we assume the homogeneous vertical structure of the disk. However, according to model calculations, protoplanetary disks have gradients in the vertical temperature and density structure \citep[e.g.][]{Chiang1997,Dullemond2002,Inoue2009}. Furthermore, observationally, \citet{Teague2018} has reported that the gas temperature $T_{gas}\approx40~K$ in the inner region ($r<100~AU$) of the TW Hya disk derived from multiple transition lines of CS molecule. Similarly, \citet{Loomis2018} also derive the rotational temperature $T_{rot}\approx32~K$ from multiple transition lines of CH$_{3}$CN molecule in the TW Hya disk. Since those molecular lines are emitted from the upper layer or surface of the disk, they indicate the vertical structure can exist in the TW Hya disk.

If the dust continuum emission is optically thin, the emission is the result of integration throughout the vertical direction of the disk, and mainly traces the disk midplane since dust grains are concentrated near the midplane by settling. Therefore, the assumption of homogeneous vertical structure would not affect the result very much. Meanwhile, if the dust continuum emission is optically thick, we need to be more careful about the vertical structure. Now, the dust opacity generally increases with frequency. That is, if the column density is high enough, the dust continuum emission at different frequencies becomes optically thick at different layers in the vertical direction of the disk. In this case, the assumption of the homogeneous vertical structure of dust temperature could affect the result of the synthetic analysis. 

If the irradiation from the central star is the dominant heating source, the dust temperature increases along the vertical direction. Figure 2 of \citet{Inoue2009} shows the temperature profile as a function of the optical depth in the vertical direction of the disk, indicating that in the region where the Planck mean optical depth with the stellar effective temperature ($T_{\ast,eff}$) is larger than one, the temperature is almost constant in the vertical direction, although it depends on the dust model. Now, the optical depths at the ALMA Bands are $\sim1-10^3$ times smaller than the Planck mean optical depth at a given $T_{*,eff}$, depending on the dust size \citep[e.g.][]{Miyake1993}. For example, the optical depths at ALMA Band 3 ($\nu=112~GHz$) and Band 10 ($\nu=870~GHz$) are $\tau_{B3}\sim0.004$ and $\tau_{B10}\sim0.03$ respectively, where the Planck mean optical depth at $T_{*,eff}$ becomes unity ($\tau_{\ast}=1$) if we adopt the dust model with the dust size of $10~\mu m$. Thus, the dust temperature is nearly constant in the region where the dust emission becomes optically thick at the ALMA Bands and it could be reasonable to assume that the effect of the dust temperature gradient on the synthetic analysis is not very large in this case. 

Meanwhile, if the viscous heating is dominant, the dust temperature increases near the midplane. In this case, the temperature gradient could affect the result of the synthetic analysis more significantly. According to model calculations \citep[e.g.][]{DAlessio1998}, the viscous heating will be dominant only in the inner region ($r < 1~AU$) if the accretion rate is small ($\dot{M}\sim10^{-8}M_{\odot} yr^{-1}$). The viscous heating becomes dominant in the outer disk for larger accretion rates \citep[e.g.][]{Dullemond2007}. 

The inclination angle of the disks could also affect the result. If the disk is almost edge-on, the dust continuum emission becomes optically thick and the assumption of a homogeneous structure along the line-of-sight is no more valid. However, in the case of the TW Hya disk, the inclination angle is small \citep[$i\sim7^{\circ}$,][]{Qi2008} and the line-of-sight direction is not very different from the vertical direction of the disk. Thus, additional observations at different frequencies will help us to constrain the dust properties when they have non-uniform structure along the line-of-sight.

\subsection{The frequency dependency of $\beta$}

We assume the dust opacity is proportional to the frequency as $\kappa_{\nu}\propto\nu^{\beta}$ where $\beta$ is independent of the frequency for the ALMA multiband analysis. This power-law index $\beta$ is strongly related to the dust grain models in the disks; the dust size distribution, the maximum/minimum size of the grains, the internal structure of the grains, and the composition of the grains. Thus, we can infer the dust properties in protoplanetary disks by comparing the observed $\beta$ with the theoretical/experimental $\beta$ derived from different conditions \citep[e.g.][]{Beckwith2000}. $\beta$ of the grains in the interstellar clouds is observationally measured as $\beta\approx 1.7$. 
Meanwhile, many observations of protoplanetary disks show $\beta\leq 1$ \citep[e.g.][]{Beckwith1991}, and recent spatially resolved observations show $\beta$ is smaller in the inner region of the disks \citep[e.g.][]{Perez2012,Tsukagoshi2016}, suggesting further grain growth close to the central stars. It is consistent with grain growth theory which leads to planet formation \citep[e.g.][]{Armitage2010}.

Theoretical/Experimental studies of dust opacity suggests that $\beta$ can vary with the frequency, depending on dust properties, such as dust temperature, grain size distribution and composition \citep[e.g.][]{Miyake1993,Mennella1998,Chihara2002,Draine2006,Koike2006,Min2016,Woitke2016}. \citet{Demyk2017a,Demyk2017b} recently reported the variations of $\beta$ on the frequencies by experimental measurements for the particles with the different dust temperatures ($T_{d}=10-300~K$) and chemical compositions (Mg-rich or Fe-rich amorphous silicates). They suggest that the $\beta$ anti-correlates with $T_{d}$ at a given frequency for $T_{d}>30~K$ due to the additional absorption processes of the grains. They also find the complex $\beta$ variations at $\lambda \geq 100\mu m$ for the different $T_{d}$ and compositions. 

Despite those theoretical/experimental studies on $\beta$ with the dust grain models, it is difficult to fully understand the frequency dependence of $\beta$ of actual grains in protoplanetary disks. Therefore, we simply assumed frequency independent $\beta$ in this work. However, we need to be cautious that $\beta$ can vary depending on the frequency. In this case, we need additional observations at different frequencies in order to constrain the dust properties.

\subsection{Extension to ALMA Band 1}

We find that frequency intervals between the bands should be large enough for better constraints on the dust properties through the multiband analysis. It infers that we can constrain better dust properties if we add lower frequency band instead of Band 3 ($112~GHz$) in the band sets. Since ALMA Band 1 receiver is developing now \citep{Huang2016}, we perform the synthetic multiband analysis to the band sets including Band 1 ($40~GHz$) in order to check the applicability of multiband analysis. 

Figure 10 presents the constraint ranges of synthetic dust properties, $T_{d,syn}(r)$, $\tau_{B1,syn}(r)$, and $\beta_{syn}(r)$, derived from five band sets including Band 1. The constraint ranges of dust properties derived from different band sets are represented by different colors; yellow for Band [9,7,1] set, purple for [9,6,1], blue for [10,7,1], red for [10,6,1], and green for [10,5,1]. The black solid lines are the model profiles of $T_{d}(r)$, $\tau_{B1,model}(r)$, and $\beta(r)$ derived by the same method described in Section \ref{subsec:model}. 

The constraint ranges of $T_{d,syn}(r)$ are better than the ones derived from the band sets including Band 3, such as Band [10,6,3] set. Similarly, the constraint ranges of $\tau_{B1,syn}(r)$ and $\beta_{syn}(r)$ become much better than ones in Figure 2 as well. We calculate the same value as in Figure 1 for the band sets with ALMA Band 1 to estimate how Band 1 improve the constraint range. Figure 11 presents the averaged normalized deviation of $T_{d,syn,max}$ (left), $\tau_{B1,syn,max}$ (middle), and $\beta_{syn,max}$ (right) from the model values for Band [x2,y2,1] sets (top-right) comparing to Band [x1,y1,3] sets (bottom-left). The deviation of $T_{d,syn,max}$ for Band [10,6,1] set is $\sim$10\%, which is twice better than the deviation for Band [10,6,3] set ($\sim$21\%). Furthermore, we have more options on band sets having $\lesssim$20\% accuracy on the constraints on dust properties, such as Band [10,8,1], [10,7,1], [10,5,1], [10,4,1], [9,6,1], and [9,5,1] sets. Since including Band 1 observation makes larger frequency intervals between the bands than Band 3, the dust properties are accurately constrained by the multiband analysis. Thus, we conclude that Band 1 observations will improve the constraints on the dust properties derived by the multiband analysis.

\begin{figure}[t]
     \centering
     \includegraphics[width=\linewidth, clip]{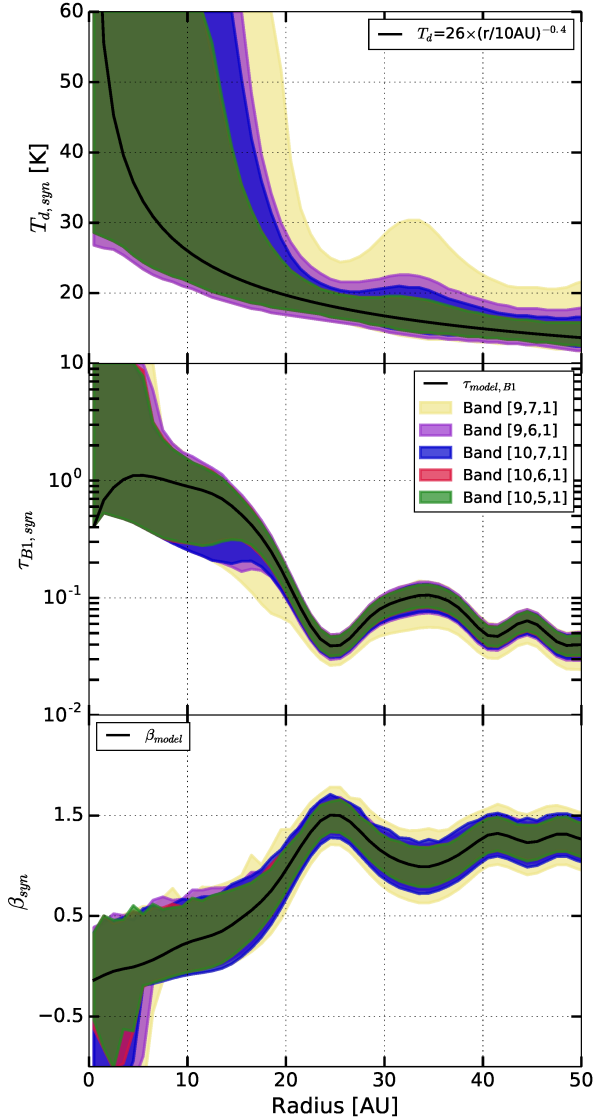}
     \caption{The constraint range of radial profiles of dust temperature $T_{d,syn}$ (top), optical depth at ALMA Band 1 $\tau_{B1,syn}$ (middle), and opacity power-law index $\beta_{syn}$ (bottom) derived from five ALMA band sets using the synthetic multiband analysis. The different colors represent the different ALMA band sets; Band [9,7,1] set by yellow, [9,6,1] by purple, [10,7,1] by blue, [10,6,1] by red, and [10,5,1] by green. The black solid line shows the model radial profile of $T_{d,model}$, $\tau_{B1,model}$, and $\beta_{model}$ derived by the same method described in Section \ref{subsec:model}. Comparing to Figure 2, the band sets including Band 1 instead of Band 3 provide better constraints on the dust properties.
     }%
     \label{fig:figure10}
\end{figure}

\begin{figure*}[t]
\centering  
\subfiguretopcaptrue
\mbox{}\hfill  
\subfigure[$T_{d,syn,max}$]{\label{fig:11a}\includegraphics[width=0.3\textwidth]{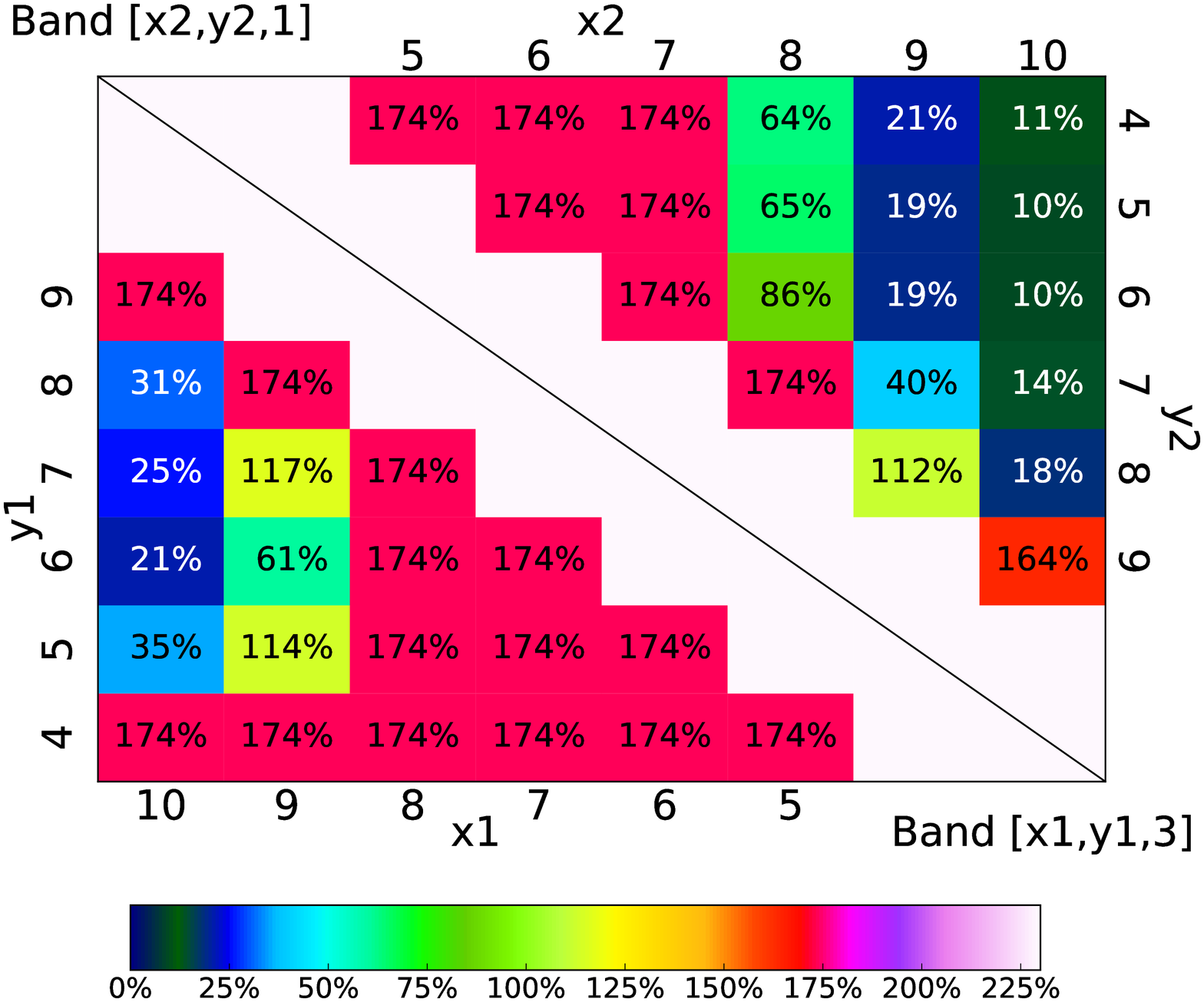}}\hfill
\subfigure[$\tau_{\nu,syn,max}$]{\label{fig:11b}\includegraphics[width=0.3\textwidth]{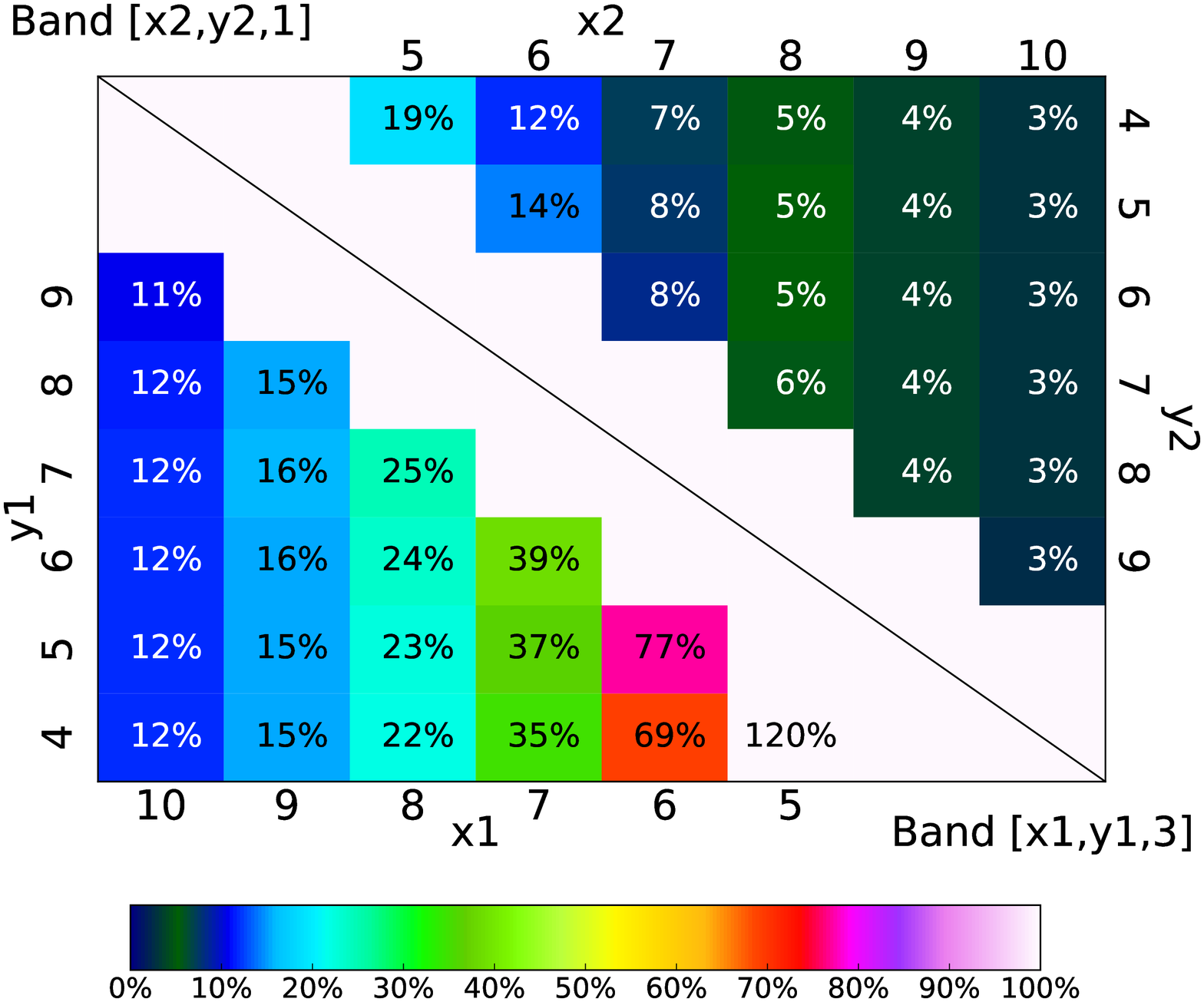}}\hfill
\subfigure[$\beta_{syn,max}$]{\label{fig:11c}\includegraphics[width=0.3\textwidth]{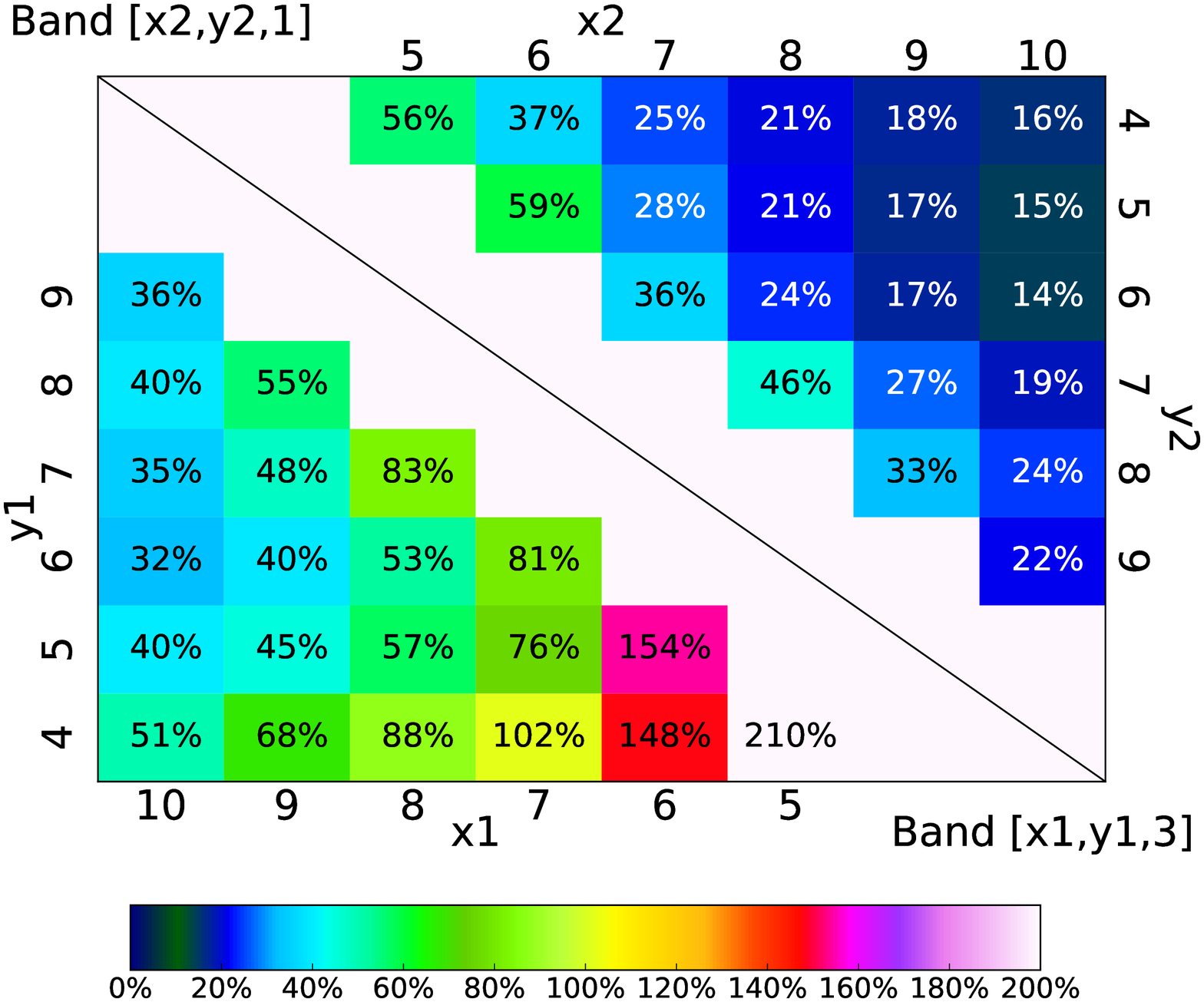}}
\hfill\mbox{}
\subfiguretopcapfalse
\caption{ The averaged normalized deviation of (a) $T_{d,syn,max}$, (b) $\tau_{\nu,synmax}$, and (c) $\beta_{syn,max}$ from the model values within $20~AU<r<45~AU$ for the Band [x1,y1,3] sets (bottom-left, $\nu$ is Band 3) and [x2,y2,1] sets (top-right, $\nu$ is Band 1). The background colors of each block indicate the averaged deviation in percentage. The deviation of $T_{d,syn,max}$ for Band [10,6,1] set is $\sim$10\% which is twice better than the one for Band [10,6,3] set, $\sim$21\%. }
\label{fig:figure11}
\end{figure*}

\section{Summary}
\label{sec:Summary}

In this paper, we try to constrain the radial profiles of dust temperature $T_{d}$, optical depth $\tau_{\nu}$ and the dust opacity power-law index $\beta$ in the TW Hya protoplanetary disk using the ALMA multiband analysis. To find which ALMA band set provides the best constraints on the dust properties, we perform the synthetic ALMA multiband analysis to all the possible combinations of three ALMA bands among Band 3 to 10 reflecting $\pm10\%$ observational errors for Band 3, 4, 5, and 6, $\pm15\%$ errors for Band 7 and 8, and $\pm20\%$ errors for Band 9 and 10.  

According to the synthetic multiband analysis results, Band [10,6,3] set gives us the best constraints of $T_{d}(r)$, $\tau_{\nu}(r)$, and $\beta(r)$ profiles. In addition, we find two conditions for the good ALMA band sets;
\begin{enumerate}
\item Band 9 or 10 should be included in the band set
\item The frequency intervals between the bands are large enough
\end{enumerate}
We examine the effect of each dust parameter on the synthetic multiband analysis results to understand why those conditions are required for the good ALMA band sets. We find that the combination of an optically thick band for the accurate $T_{d}$ estimation and optically thin band for the accurate $\tau_{\nu}$ estimation is essential for the good constraints on the dust properties. We also find that large $\beta$ leads better constraints on the dust properties. When $\beta$ becomes larger, the high-frequency band becomes optically thick and low-frequency band becomes optically thin. Therefore, we can obtain the combination of optically thick and optically thin band. The blackbody curves at low dust temperature deviate from the Rayleigh-Jeans (RJ) limit at high frequency. Band 9 or 10 has an advantage on the $T_{d}$ estimation due to the large deviation of the blackbody curves from RJ limit. In addition, the large frequency intervals between the bands reduce the error of spectral index $\alpha$. Accurate measurement of $\alpha$ leads to good constraints on the dust properties, $T_{d}$, $\tau_{\nu}$ and $\beta$. We also examine the possibility to apply the multiband analysis to various disk conditions using the Monte-Carlo method. For various model dust properties, we obtain good constraint on dust properties with low $T_{d,model}$, intermediate $\tau_{190GHz,model}$ and high $\beta_{model}$. That is, we can apply the multiband analysis to not only the TW Hya disk but the general protoplanetary disks.

To check the synthetic multiband analysis is consistent with the observational data, we apply the multiband analysis to the ALMA archival data of the TW Hya disk at Band 4, 6, 7 and 9. First, we apply the multiband analysis to the high-resolution dust continuum images at Band 7, 6, and 4. The derived $T_{d}(r)$ profile is smaller than the $T_{d,model}(r)$ profile and has a bump around $r\sim33~AU$ which is probably caused by the large uncertainty on constraint range due to the $\beta$ drop at the same location. The constraint range of $T_{d}$ corresponding to the observational errors derived from the Band [7,6,4] set is too broad to specify $T_{d}$ value at every radius. The derived $\tau_{B4}(r)$ and $\beta(r)$ profiles follow the small-scale structures of the model profiles, especially $r>35~AU$, but the constraint ranges corresponding to the observational errors are broad like the $T_{d}$ constraint range. It is consistent with the synthetic multiband analysis result that the dust properties derived from Band [7,6,4] data are sensitive to the observational errors.

We also apply the same analysis to the low-resolution ALMA Band 9 data with the smoothed Band 6 and 4 data. The derived $T_{d}(r)$ profile shows a smooth profile close to the model profile in $r \leq 40~AU$. The derived $\tau_{B4}(r)$ and $\beta(r)$ profiles also follow the model profiles very well in the same region. Although the constraint range of $T_{d}$ corresponding to the observational errors derived from Band [9,6,4] set is still broad to specify the accurate values, the constraint ranges of $\tau_{B4}$ and $\beta$ become narrower than the ranges derived from Band [7,6,4] set. It is consistent with the synthetic multiband analysis result that the dust properties derived from Band [9,6,4] data are less affected by the observational errors than those derived from Band [7,6,4] data because Band [9,6,4] set satisfies the two conditions for the good band sets. Unfortunately, the large beam size of Band 9 smoothes out the small-scale structures in the radial profiles of the dust properties during the convolution process of the images. Therefore, we need observations with a high spatial resolution dust continuum observation at Band 9 or 10 to figure out the dust properties in the gap structures at $\sim25~AU$ and $\sim40~AU$ in the TW Hya disk.

The multiband analysis is a good way to constrain the dust properties in the protoplanetary disks from the observational data. However, we still have some points to improve the multiband analysis. First, we assume the vertical structure of the disk is homogeneous in the calculation but the disks actually have the gradient of temperature and density profiles along the vertical direction. We need to consider those vertical disk structures to obtain better constraints of the dust properties. Another point is that the frequency dependence of $\beta$ is not taken into account in the calculation while the theoretical/experimental researches have reported the possible dependence of $\beta$ on the frequency and temperature. Thus, we should be cautious that $\beta$ could depend on the frequency. We will need more than three bands to constrain the dust properties if the vertical structure of the dust properties is significant and/or $\beta$ has a strong dependence on the frequency.

\acknowledgments

We thank the referee for his/her constructive comments. This paper makes use of the following ALMA data: ADS/ JAO.ALMA\#2012.1.00422.S, 2015.1.00005.S, and 2015.1.00686.S. Part of the data are retrieved from the JVO portal (http://jvo.nao.ac.jp/portal/alma/archive.do) operated by the NAOJ. Data analysis was carried out on the Multi-wavelength Data Analysis System operated by the Astronomy Data Center (ADC), National Astronomical Observatory of Japan. ALMA is a partnership of ESO (representing its member states), NSF (U.S.) and NINS (Japan), together with NRC (Canada) and NSC and ASIAA (Taiwan) and KASI (Republic of Korea), in cooperation with the Republic of Chile. The joint ALMA observatory is operated by ESO, AUI/NRAO, and NAOJ. Seongjoong Kim was supported by the ALMA Japan Research Grant of NAOJ Chile Observatory, NAOJ-ALMA-199. Takashi Tsukagoshi was supported by JSPS KAKENHI grant number 17K14244. This work is supported by MEXT Grants-in-Aid for Scientific Research 18H05441.

\bibliographystyle{apj}
\bibliography{../ALL_papers.bib}

\end{document}